\documentclass[aps,pra,onecolumn,amssymb,amsfonts,floatfix,nofootinbib,superscriptaddress,longbibliography,notitlepage]{revtex4-1}
\pdfoutput=1
\usepackage[utf8]{inputenc}
\usepackage{color}
\usepackage{multirow}
\usepackage{dcolumn}
\usepackage{mathbbol}
\usepackage{xcolor}
\usepackage{amsmath,amsthm,amsfonts,amssymb}
\usepackage{times}
\usepackage[normalem]{ulem}
\usepackage{physics}
\usepackage{commath}
\usepackage{bm}
\usepackage[pdftex]{graphicx}
\usepackage[colorlinks=false]{hyperref}

\newcommand{\flow}[1] {
  {\bm \varphi}^{#1}}

\begin{document}

\title[Quantum vs classical dynamics in a spin-boson system]{Quantum vs classical dynamics in a spin-boson system: \\
manifestations of spectral correlations and scarring}

\author{D Villase\~nor} 
\author{S Pilatowsky-Cameo}
\address{Instituto de Ciencias Nucleares, Universidad Nacional Aut\'onoma de M\'exico, Apdo. Postal 70-543, C.P. 04510  Cd. Mx., Mexico}
\author{M~A~Bastarrachea-Magnani}
\address{Department of Physics and Astronomy, Aarhus University, Ny Munkegade, DK-8000 Aarhus C, Denmark}
\author{S~Lerma-Hern\'andez}
\address{Facultad de F\'isica, Universidad Veracruzana, Circuito Aguirre Beltr\'an s/n, Xalapa, Veracruz 91000, Mexico}
\author{L~F~Santos}
\address{Department of Physics, Yeshiva University, New York, New York 10016, USA}
\author{J G Hirsch}
\address{Instituto de Ciencias Nucleares, Universidad Nacional Aut\'onoma de M\'exico, Apdo. Postal 70-543, C.P. 04510  Cd. Mx., Mexico}
%%%%%%%%%%%%%% ABSTRACT %%%%%%%%%%%%%%%%
   
\begin{abstract}
We compare the entire classical and quantum evolutions of the Dicke model in its regular and chaotic domains. This is a paradigmatic interacting spin-boson model of great experimental interest. By studying the classical and quantum survival probabilities of initial coherent states, we identify features of the long-time dynamics that are purely quantum and discuss their impact on the equilibration times. We show that the ratio between the quantum and classical asymptotic values of the survival probability serves as a metric to determine the proximity to a separatrix in the regular regime and to distinguish between two manifestations of quantum chaos: scarring and ergodicity. In the case of maximal quantum ergodicity, our results are analytical and show that quantum equilibration takes longer than classical equilibration. 
\end{abstract}

%\keywords{Quantum chaos, quantum dynamics, quantum ergodicity, scarring}
%\submitto{\NJP}

\maketitle
%%%%%%%%%%%%%%%%%%%%%%%%%%%%%%%%%%%%%%%%

%%%%%%%%%%%%%%%%%%%%%%%%%%%%%%%%%%%%%%%%%%%%%%%%%%%%%%%
%%%%%%%%%%%%%% SEC.1  INTRODUCTION %%%%%%%%%%%%%%%%
%%%%%%%%%%%%%%%%%%%%%%%%%%%%%%%%%%%%%%%%%%%%%%%%%%%%%%%
\section{INTRODUCTION}
\label{sec:1}

Experimental advances in the studies of isolated quantum systems have resulted in ever longer coherence times~\cite{kinoshita06,Simon2011,Schreiber2015,Kaufman2016}, prompting theoretical and experimental analysis of long-time quantum dynamics. Recent works on quantum chaos have shown that the short-time exponential growth of out-of-time-ordered correlators (OTOCs) \cite{Maldacena2016PRD,Rozenbaum2017,Rozenbaum2019,Hashimoto2017,Garcia2018,Jalabert2018,Chavez2019} is not a universal signature of chaos, but can emerge also near critical points~\cite{Pappalardi2018,Hummel2019,Pilatowsky2019}. This has offered another motivation to switch the attention to long-time dynamics~\cite{Hummel2019,Fortes2019}. 

At large times, equilibration eventually occurs, and a main question in studies of nonequilibrium quantum dynamics is how long it takes for this to happen. In classical mechanics, the mixing properties of chaotic dynamics have provided a fundamental mechanism to explain the equilibration process as well as the ergodic properties of physical systems. In quantum mechanics, even though isolated systems are described by linear equations, one can still talk about equilibration in the sense of saturation of the dynamics, that is, the evolution of observables reaches a point where it simply fluctuates around its asymptotic value, while these fluctuations decrease with the system size~\cite{Reimann2008,Short2011,Short2012,Zangara2013,HeSantos2013}. Since the passage from the classical to the quantum domain entails new phenomena, such as superpositions, quantum interferences, the effects of universal spectral correlations~\cite{Leviandier1986,Wilkie1991,Alhassid1992,Torres2017PTR,Torres2018,Schiulaz2019,Lerma2019,Cotler2017,Numasawa2019}, and quantum scars~\cite{Pechukas1982,Stechel1984,Heller1984,Heller1987,Heller1991,Bies2001,Berry1977,Kay1983,Percival1973,Turner2018,Ho2019}, one may ask what differences and similarities one should find between the classical and quantum approaches to equilibrium. As we discuss here, the quantum-classical correspondence breaks down at long times, when effects that are purely quantum come to light.

The present work investigates the entire quantum and classical evolutions of the Dicke model, with the purpose of identifying general features. This two-degree-of-freedom interacting spin-boson model was introduced to explain the collective phenomenon of superradiance~\cite{Dicke1954,Hepp1973a,Hepp1973b,Wang1973,Emary2003,Garraway2011}, a phenomenon that has been experimentally studied with cold atoms in optical cavities~\cite{Baumann2010,Baumann2011,Ritsch2013,Baden2014,Klinder2015,Kollar2017}. Depending on the parameters and excitation energy, the model presents regular and chaotic domains~\cite{Lewenkopf1991,Emary2003PRL,Emary2003,Bastarrachea2014b,Bastarrachea2015,Bastarrachea2016PRE,Chavez2016}. It has been used in studies of nonequilibrium dynamics~\cite{Fernandez2011,Altland2012PRL,Shen2017,Lerma2019,Kloc2018,Kirton2019} and as a paradigm of the ultra-strong coupling regime in several systems~\cite{DeBernardis2018,Kockum2019,FornDiaz2019}. Experimentally, the model can be studied by means of cavity assisted Raman transitions~\cite{Baden2014,Zhang2018} and with trapped ions~\cite{Cohn2018,Safavi2018}.

By comparing the entire quantum and classical evolutions, we  provide a broad picture of the dynamics and find when and why the quantum-classical correspondence no longer holds. We place initial coherent states in the regular and chaotic regions of the model and study the probability to find the initial state at a later time, the so-called survival probability or return probability. We choose coherent states, because they enable a direct comparison between the exact quantum evolution and its classical description obtained with the truncated Wigner approximation (TWA)~\cite{Steel1998,Polkovnikov2010,Schachenmayer2015}. An advantage of using the survival probability is that it recognizes differences in the quantum and classical dynamics that are not limited to the quantum fluctuations after saturation. 

Results are presented for various initial states and four representative ones are studied at length. They are selected according to their level of delocalization in the energy eigenbasis. At short times, the quantum and classical evolutions coincide. At long times, four general cases are singled out, as listed below. They are distinguished according to the proximity to a separatrix in the regular regime and according to the manifestations of quantum chaos that are observed: scarring or ergodicity.

(i) In the regular regime, if the initial state is far from a separatrix, the quantum and classical evolutions coincide up to equilibration. Beyond this point, the classical survival probability reaches a constant value, while the quantum survival probability fluctuates around the classical asymptotic value.

(ii) In the regular regime, if the initial state is close to a separatrix, small differences between the quantum and classical evolutions appear at long times due to a tunneling effect that exists only in the quantum domain.

(iii) In the chaotic region, the quantum survival probability equilibrates faster than the classical one, when the quantum initial state has large components in scarred eigenstates. Quantum scars are signatures in the Hilbert space of the presence of classical unstable periodic orbits.

(iv) In the chaotic region, if the initial state is highly delocalized in the energy eigenbasis (ergodic), the quantum survival probability takes longer to equilibrate than its classical counterpart. This is because the quantum survival probability equilibrates only after passing through the correlation hole, which is a dynamical manifestation of spectral correlations that is nonexistent in the classical limit.

We show that the ratio between the asymptotic values of the quantum and classical survival probabilities can be used as a metric to distinguish the four cases above.  A ratio equal to two indicates maximal quantum ergodicity, as in (iv). For this case, we derive an analytical expression for the evolution of the quantum survival probability and its equilibration time. 

We also compare the results for the survival probability with the classical evolution of the Wigner distribution in phase space. With this parallel, we gain a deeper understanding of specific features of the quantum dynamics that emerge at different time scales.

This paper is organized as follows. The core of the work is in Sec.~\ref{sec:4}, where a detailed comparative study of the classical and quantum evolution of the survival probability is presented for the regular and chaotic regimes. In preparation for this analysis, Sec.~\ref{sec:2} describes the Dicke Hamiltonian and its classical limit, and Sec.~\ref{sec:QuanDyn} describes the initial coherent states studied. In Sec.~\ref{sec:3}, we introduce the survival probability, its relation with the local density of states (LDoS), and its classical approximation using the TWA. Our conclusions are presented in Sec.~\ref{sec:5}.

%%%%%%%%%%%%%%%%%%%%%%%%%%%%%%%%%%%%%%%%%%%%%%%%%%%%%%%
%%%%%%%%%%%%%% SEC.2  QUANTUM HAMILTONIAN AND MODEL %%%%%%%%%%%%%%%%
%%%%%%%%%%%%%%%%%%%%%%%%%%%%%%%%%%%%%%%%%%%%%%%%%%%%%%%
\section{DICKE MODEL}  
\label{sec:2}

The Dicke model~\cite{Dicke1954} represents a set of $\mathcal{N}$ two-level atoms with atomic transition frequency $\omega_{0}$ interacting with a single mode of a radiation field with frequency $\omega$. It is described by the following Hamiltonian,
\begin{equation}
\label{eqn:qua_hamiltonian}
\hat{H}_{D}=\omega\hat{a}^{\dagger}\hat{a}+\omega_{0}\hat{J}_{z}+\frac{2\gamma}{\sqrt{\mathcal{N}}}\hat{J}_{x}(\hat{a}^{\dagger}+\hat{a}),
\end{equation}
where $\hbar=1$, $\hat{a}$ ($\hat{a}^{\dagger}$) is the bosonic annihilation (creation) operator of the field mode, $\hat{J}_{x,y,z}=\frac{1}{2}\sum_{k=1}^{\mathcal{N}}\hat{\sigma}_{x,y,z}^{k}$ are collective pseudo-spin operators given by the sum of the Pauli matrices $\hat{\sigma}_{x,y,z}$, and $\gamma$ is the spin-boson interaction strength. When $\gamma$ reaches a critical value $\gamma_{c}=\sqrt{\omega\omega_{0}}/2$, a second-order quantum phase transition takes place in the system~\cite{Hepp1973a,Hepp1973b,Wang1973,Emary2003}. It goes from the normal phase ($\gamma<\gamma_c$), where the ground state is characterized by all atoms in their ground state and no photons, to the the superradiant phase ($\gamma>\gamma_c$), where the ground state has a macroscopic population of photons and excited atoms.

The eigenvalues $j(j+1)$ of the total spin operator $\hat{\textbf{J}}^{2}=\hat{J}_{x}^{2}+\hat{J}_{y}^{2}+\hat{J}_{z}^{2}$ determine the different invariant subspaces. We work with the maximum value $j=\mathcal{N}/2$, which defines a symmetric atomic subspace that includes the ground state. The Hamiltonian $\hat{H}_{D}$ commutes also with the parity operator $\hat{\Pi}=e^{i\pi\hat{\Lambda}}$, where $\hat{\Lambda}=\hat{a}^{\dagger}\hat{a}+\hat{J}_{z}+j\hat{1}$ represents the total number of excitations with eigenvalues $\Lambda=n+m+j$. Here, $n$ indicates the number of photons and $m+j$ is the number of excited atoms,  $m$ being the eigenvalue of the operator $\hat{J}_z$.

%%%%%%%%%%%%%%%%%%%%%%%%%%%%%%%%%%%%%%%%%%%%%%%%%%%%%%%
%%%%%%%%%%%%%%%%%%%%%%%%%%%%%%%%%%%%%%%%%%%%%%%%%%%%%%%
\subsection{Classical Limit}

The corresponding classical Hamiltonian is obtained using Glauber coherent states for the bosonic sector~\cite{Deaguiar1991,Deaguiar1992,Bastarrachea2014a,Bastarrachea2014b,Bastarrachea2015,Chavez2016},
\begin{equation}
\label{eqn:glauber}
|q,p\rangle=e^{-(j/4)\left(q^{2}+p^{2}\right)}e^{\left[\sqrt{j/2}\left(q+ip\right)\right]\hat{a}^{\dagger}}|0\rangle, 
\end{equation}
and Bloch coherent states for the pseudo-spin sector, 
\begin{equation}
\label{eqn:bloch}
|Q,P\rangle=\left(1-\frac{Z^{2}}{4}\right)^{j}e^{\left[\left(Q+iP\right)/\sqrt{4-Z^{2}}\right]\hat{J}_{+}}|j,-j\rangle,
\end{equation}
where $Z^{2}=Q^{2}+P^{2}$. The canonical $j$-independent variables $(q,p)$ and $(Q,P)$ are associated with the photonic and atomic degrees of freedom, respectively. The state $|0\rangle$ denotes the photon vacuum and $|j,-j\rangle$, the state with all atoms in their ground state. The rescaled classical Hamiltonian $h_{cl}$ is given by (see \ref{app:1} for details),
\begin{align}
\label{eqn:cla_hamiltonain_QP}
h_{cl} & \equiv \frac{\langle q,p;Q,P|\hat{H}_{D}|q,p;Q,P\rangle}{j}\\
&=\frac{\omega}{2}(q^{2}+p^{2})
 +\frac{\omega_{0}}{2}Z^{2} +2\gamma Q q\sqrt{1-\frac{Z^{2}}{4}} -\omega_{0}.\nonumber
\end{align}
The rescaled classical Hamiltonian and its four-dimensional phase space $\mathcal{M}$ are independent of $j$. This is equivalent to working with an effective Planck constant $\hbar_{\text{eff}}=1/j$~\cite{Ribeiro2006}. 

\subsection{Hamiltonian Parameters}

The Hamiltonian parameters $\omega$, $\omega_0$, and $\gamma$ are chosen, so that the regular and chaotic regimes are clearly identified. The parameter $\gamma$ controls the coupling between photons and atoms in the system and therefore the emergence of a chaotic region. We select the coupling strength in the superradiant phase, $\gamma=2\gamma_c$, where the presence of chaotic behavior is guaranteed~\cite{Chavez2016}. We choose $\omega=\omega_0$ for convenience, but other choices would not change the broad picture developed in this work.

For the system size $j=100$ considered here, the normalized energy
\begin{equation}
\label{eqn:norm_ener}
\epsilon=\frac{E}{\omega_{0}j}
\end{equation}
of the ground state is $\epsilon_{GS}=-2.125$. The dynamics is regular from $\epsilon_{GS}$ up to $\epsilon\approx-1.7$ and chaotic above this point~\cite{Chavez2016}.

%%%%%%%%%%%%%%%%%%%%%%%%%%%%%%%%%%%%%%%%%%%%%%%%%%%%%%%
%%%%%%%%%%%%%%%%%%%%%%%%%%%%%%%%%%%%%%%%%%%%%%%%%%%%%%%
\subsection{Numerical Diagonalization}

The effective Planck's constant $1/j$ determines the resolution that a coherent initial state has in the classical phase space. This means that a large $j$ is necessary for the quantum dynamics to reflect the classical effects and for the purely quantum properties to be identified. Large systems are needed to resolve the structure of the classical phase space~\cite{Bakemeier2013}. We are able to consider a large system size, $j=100$, because we employ an efficient basis that guarantees the convergence of the eigenvalues and eigenstates for a broad part of the bounded spectrum of the Dicke model~\cite{Bastarrachea2014PSa,Bastarrachea2014PSb}. The dimension of the truncated Hilbert space in this efficient basis is given by $\text{dim}=(2j+1)(N_{\text{max}}+1)$, where $N_{\text{max}}$ is an upper bound to the modified bosonic subspace. For our case study, $N_{\text{max}}=300$, so the dimension of the truncated Hilbert space is $\text{dim}=60\,501$. This ensures 
\begin{equation}
N_{c}=30\,825
\end{equation}
converged eigenstates and eigenenergies, which range from the ground state energy $\epsilon_{GS}=-2.125$ up to a truncation energy $\epsilon_{T}=0.853$.

%%%%%%%%%%%%%%%%%%%%%%%%%%%%%%%%%%%%%%%%%%%%%%%%%%%%%%%
%%%%%%%%%%%%%%%%%%%%%%%%%%%%%%%%%%%%%%%%%%%%%%%%%%%%%%%
\section{INITIAL STATES}
\label{sec:QuanDyn}

The selected initial states in this work are the Glauber-Bloch coherent states $|q,p;Q,P\rangle$. They allow for a direct connection between the quantum states and the coordinates $(q,p,Q,P)$ in the classical phase space of $h_{cl}$, and also for a relatively simple calculation of the Wigner distribution needed for the evaluation of the classical dynamics. To perform the numerical quantum calculations, the coherent states $|q,p;Q,P\rangle$ are expanded in the energy eigenstates through the efficient basis~\cite{Bastarrachea2016PRE}.

To choose the initial states, we restrict ourselves to the hyperplane $p=0$ and solve the second-degree equation $h_{cl} (q,p,Q,P)=\omega_{0}\epsilon$ in $q$. This equation has two solutions, $q_{-}$ and $q_{+}$, with $q_{-}\leq q_{+}$. The four representative initial states that we use are centered at  $(q,p,Q,P)=(q_{+},0,Q_{0},P_{0})$. Two of them have energy in the regular region, $\epsilon_{R}=-1.8$, and the other two have the energy shell fully covered by chaotic trajectories, $\epsilon_{C}=-0.5$. The criteria for our choices and the specific values of the coordinates $(Q_{0},P_{0})$ are described below. 
In Sec.~\ref{sec:4}, we provide a detailed analysis of the classical and quantum evolution of these four representative initial states. The results are then confirmed for several other initial states in the regular (see Fig.~\ref{fig:6}) and chaotic region (see Fig.~\ref{fig:9}).

%%%%%%%%%%%%%%%%%%%%%%%%%%%%%%%%%%%%%%%%%%%%%%%%%%%%%%%
\subsection{Regular Regime: $\epsilon_{R}=-1.8$}

\begin{figure}[h!]
\centering{
\includegraphics[width=10cm]{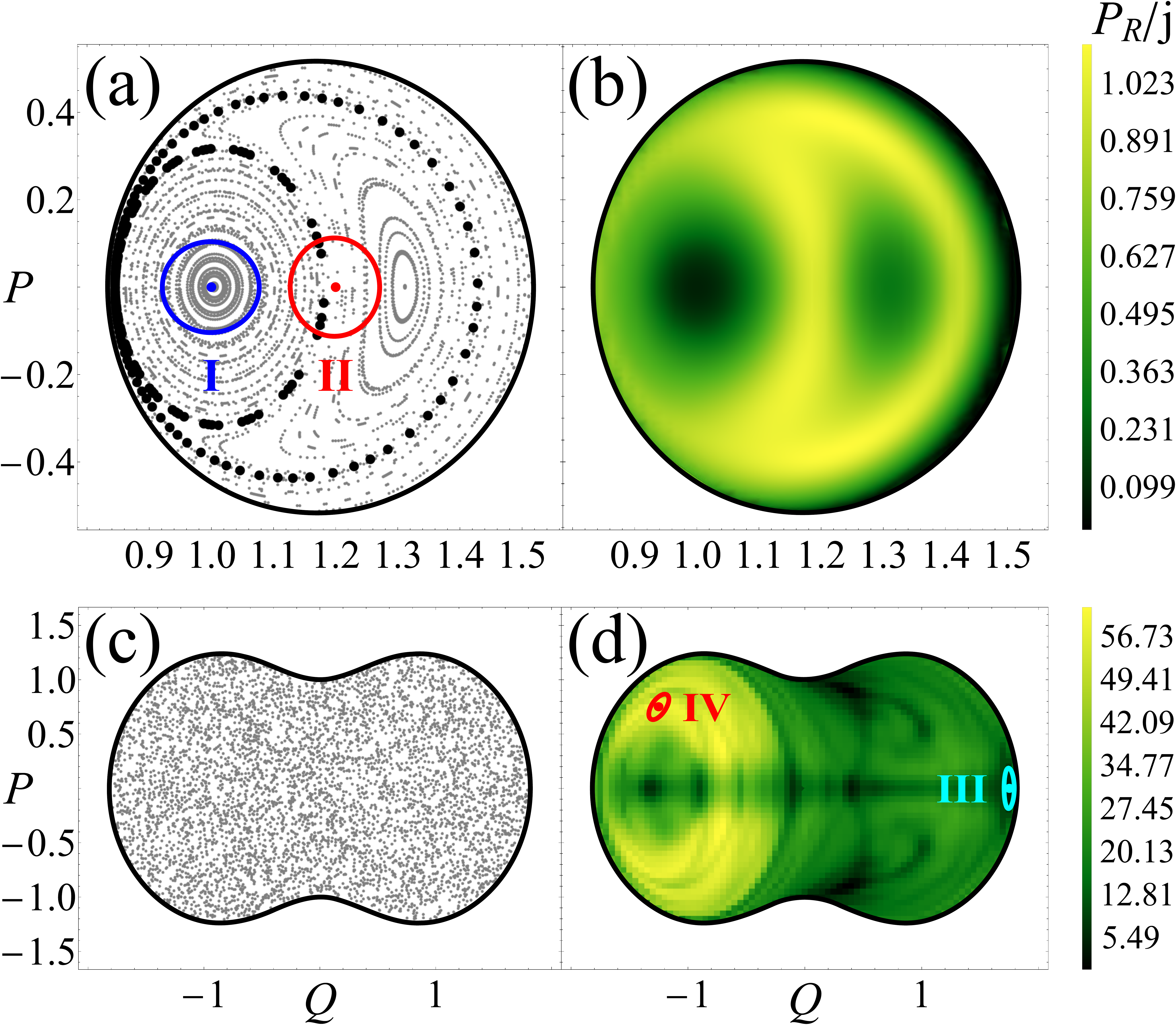}
}
\caption{Poincar\'e sections ($p=0$) for the rescaled classical Hamiltonian $h_{cl}$ at energies $\epsilon_{R}=-1.8$ (a) and $\epsilon_{C}=-0.5$ (c). 
In panel (a), large black points mark the separatrix of the regular modes, the blue point indicates the center of the coherent state I with coordinates $(Q_{0},P_{0})=(1,0)$, and the red point indicates the state II centered close to the separatrix at $(Q_{0},P_{0})=(1.2,0)$. The closed curves encircling these two points represent the spreading of the coherent state wave function up to $e^{-1}$. 
In (b) and (d): Participation ratio [Eq.~(\ref{Eq:PR})] of the coherent states centered in each point of the Poincar\'e surface and projected in the Hamiltonian eigenbasis.
 In panel (d), the cyan point indicates the initial coherent state III centered at $(Q_{0},P_{0})=(1.75,0)$, which has a low PR ($P_{R}=1066$) and the red point marks the initial state IV centered at $(Q_{0},P_{0})=(-1.25,0.75)$, which has a high PR ($P_{R}=5743$).
}
\label{fig:1}
\end{figure}

Poincar\'e sections at $p=0$ of the classical Hamiltonian at energy $\epsilon_{R}=-1.8$ are shown in Fig.~\ref{fig:1}~(a). The phase space splits in three regions of regular trajectories, whose borders are defined by the separatrix marked with large black dots. We choose state I at $(Q_{0},P_{0})=(1,0)$ and indicate it with a blue dot in Fig.~\ref{fig:1}~(a). State II is close to the separatrix, $(Q_{0},P_{0})=(1.2,0)$, and is shown with a red dot in the same figure.

%%%%%%%%%%%%%%%%%%%%%%%%%%%%%%%%%%%%%%%%%%%%%%%%%%%%%%%
%%%%%%%%%%%%%%%%%%%%%%%%%%%%%%%%%%%%%%%%%%%%%%%%%%%%%%%      

The structure of the phase space reflects the quasi-conserved quantities of the Dicke Hamiltonian at low energies~\cite{Relano2016}. They can be identified by means of an adiabatic approximation, where the dynamics separates into two parts, one fast-evolving mode and a slow one, effectively decoupling the boson and pseudo-spin dynamics~\cite{Relano2016EPL,Bastarrachea2017JPA}. For the parameters considered here, a quasi-constant of motion is given by the nutation angle of the pseudo-spin, which precesses fast around an axis whose direction oscillates slowly in a way dictated by the slow bosonic variables. State I is at the center of the slow-boson regular region located between $0.85\leq Q\leq1.16$, where the boson modes are expressed by very small nutation angles and large amplitudes of the precession axis' oscillations. The rightmost region in Fig.~\ref{fig:1}~(a) corresponds to the mode of the fast pseudo-spin degree of freedom, where the dynamics has maximal nutation angles and very small oscillations of the precession axis instead. 

Quasi-integral behaviors are destroyed in non-linear systems due to the growth and consequent overlap of nonlinear resonances between the integrable modes. These non-linear resonances arise locally in the phase space~\cite{Reichl1987} and  are separated by a separatrix, around which a stochastic layer is formed~\cite{Zaslavsky1991}. Chaos emerges from it, destroying the remnant regular surfaces  and opening the phase space to the diffusion of trajectories, which start wandering through the whole region of the resonance overlap. In the Poincar\'e sections of Fig.~\ref{fig:1}~(a), one can see non-linear resonances between the adiabatic modes described above, specifically in the moon-shaped region of regular trajectories rotating around the point $(Q,P)\simeq(1.3,0)$. State II is in this region, but very close to the separatrix.

In Fig.~\ref{fig:1}~(b), we plot the participation ratio $P_R$ of the coherent states centered in each point of the Poincar\'e surface and projected into the energy eigenstates,
\begin{equation}
P_R = \frac{1}{\sum_k |c_k|^4},
\label{Eq:PR}
\end{equation}
where $c_k = \langle E_k|q_{+},0;Q_{0},P_{0} \rangle$ and $\hat{H}|E_{k}\rangle=E_{k}| E_k\rangle$. This quantity measures the level of delocalization of the initial state in the energy eigenbasis. It varies from the minimum value, $P_{R}=1$, when the initial state coincides with an eigenstate and is therefore maximally localized, to the largest value, $P_{R}=N_c$, when all eigenstates participate equally in the evolution of the initial state. In the case of a random vector, when the components $c_k$ are uncorrelated random numbers from a Gaussian distribution, $P_{R}=N_c/3$.

The participation ratio is strongly correlated with the underlying classical dynamics~\cite{Bastarrachea2016PRE,Bastarrachea2017}. For instance, the stochastic layer appearing in the classical dynamics around the separatrix in Fig.~\ref{fig:1}~(a) is associated with large values of $P_R$ in Fig.~\ref{fig:1}~(b).  The two initial states that we chose in the regular regime have the lowest and  largest $P_R$ for that energy surface. These extremal cases allow us to identify the fundamental mechanisms underlying the evolution of  initial coherent states.

%%%%%%%%%%%%%%%%%%%%%%%%%%%%%%%%%%%%%%%%%%%%%%%%%%%%%%%
\subsection{Chaotic Regime: $\epsilon_{C}=-0.5$}

Poincar\'e sections for the selected energy $\epsilon_{C}=-0.5$ are shown in Fig.~\ref{fig:1}~(c). They reveal a region of hard chaos, where all the chaotic trajectories have the same positive Lyapunov exponent and densely fill the whole phase space. From these Poincar\'e sections, no particular region can be identified, but the participation-ratio map in Fig.~\ref{fig:1}~(d) provides a richer picture, with coherent states showing different levels of spreading  in the energy eigenbasis. To analyze how the structure of the initial states affects the dynamics and equilibration, we select two initial states. State III, indicated with the cyan point in Fig.~\ref{fig:1}~(d), is located in the region with small values of $P_R$, at $(Q_{0},P_{0})=(1.75,0)$. State IV, marked with a red point in Fig.~\ref{fig:1}~(d), is in the region of large values of $P_R$, at $(Q_{0},P_{0})=(-1.25,0.75)$.  As we show in Sec.~\ref{sec:4}, these two states are representative of the typical quantum dynamics found in chaotic regions.

%%%%%%%%%%%%%%%%%%%%%%%%%%%%%%%%%%%%%%%%%%%%%%%%%%%%%%%
%%%%%%%%%%%% SEC.3  SURVIIIAL PROBABILITY %%%%%%%%%%%%%
%%%%%%%%%%%%%%%%%%%%%%%%%%%%%%%%%%%%%%%%%%%%%%%%%%%%%%%
\section{SURVIVAL PROBABILITY}
\label{sec:3}

The survival probability, $S_{P}$, is the probability of finding an evolved quantum state back in the initial state $|\Psi(0)\rangle=\sum_{k}c_{k}| E_k\rangle$,
\begin{equation}
\label{eqn:qua_sp}
S_{P}(t)=|\langle\Psi(0)|\Psi(t)\rangle|^{2}=\left|\sum_{k}|c_{k}|^{2}e^{-i E_{k}t}\right|^{2}.
\end{equation}
By introducing the local density of states (LDoS) or strength function, that is the energy distribution weighted by the components $|c_{k}|^{2}$ of the initial state,
\begin{equation}
\label{eqn:ldos}
\mathcal{G}(E)=\sum_k |c_{k}|^{2} \delta(E-E_k),
\end{equation}
we can also write the survival probability as the squared norm of the Fourier transform of $\mathcal{G}(E)$ %as
\begin{equation}
S_{P}(t)=\left|\int \dif E\, \mathcal{G}(E) e^{-i E t}\right|^2.
\label{eq:SPFourier}
\end{equation}

The evolution of the survival probability shows different behaviors at different time scales~\cite{Tavora2016,Tavora2017,Torres2018,Schiulaz2019}. By smoothing the LDoS, one gets insight on how its structure affects the dynamics at different times. The smoothing is done through a finite resolution function, given by
\begin{equation}
\label{eqn:ldos_res}
\rho_{T}(E)=\sum_{k}|c_{k}|^{2}\Upsilon_{T}(E-E_{k}),
\end{equation}
where $\Upsilon_{T}(E-E_{k})=(T/\pi)\text{sinc}[(E-E_{k})T]$, and $\text{sinc}(x)=\sin(x)/x$. The time resolution $T =\pi/\Delta$ reflects aspects of the LDoS that are of order $\Delta$ in energy.

%%%%%%%%%%%%%%%%%%%%%%%%%%%%%%%%%%%%%%%%%%%%%%%%%%%%%%%
%%%%%%%%%%%%%%%%%%%%%%%%%%%%%%%%%%%%%%%%%%%%%%%%%%%%%%%
\subsection{Time Scales of the Survival Probability}
\label{subsect:TimeScales}

\begin{figure}[h!]
\centering{
\includegraphics[width=11cm]{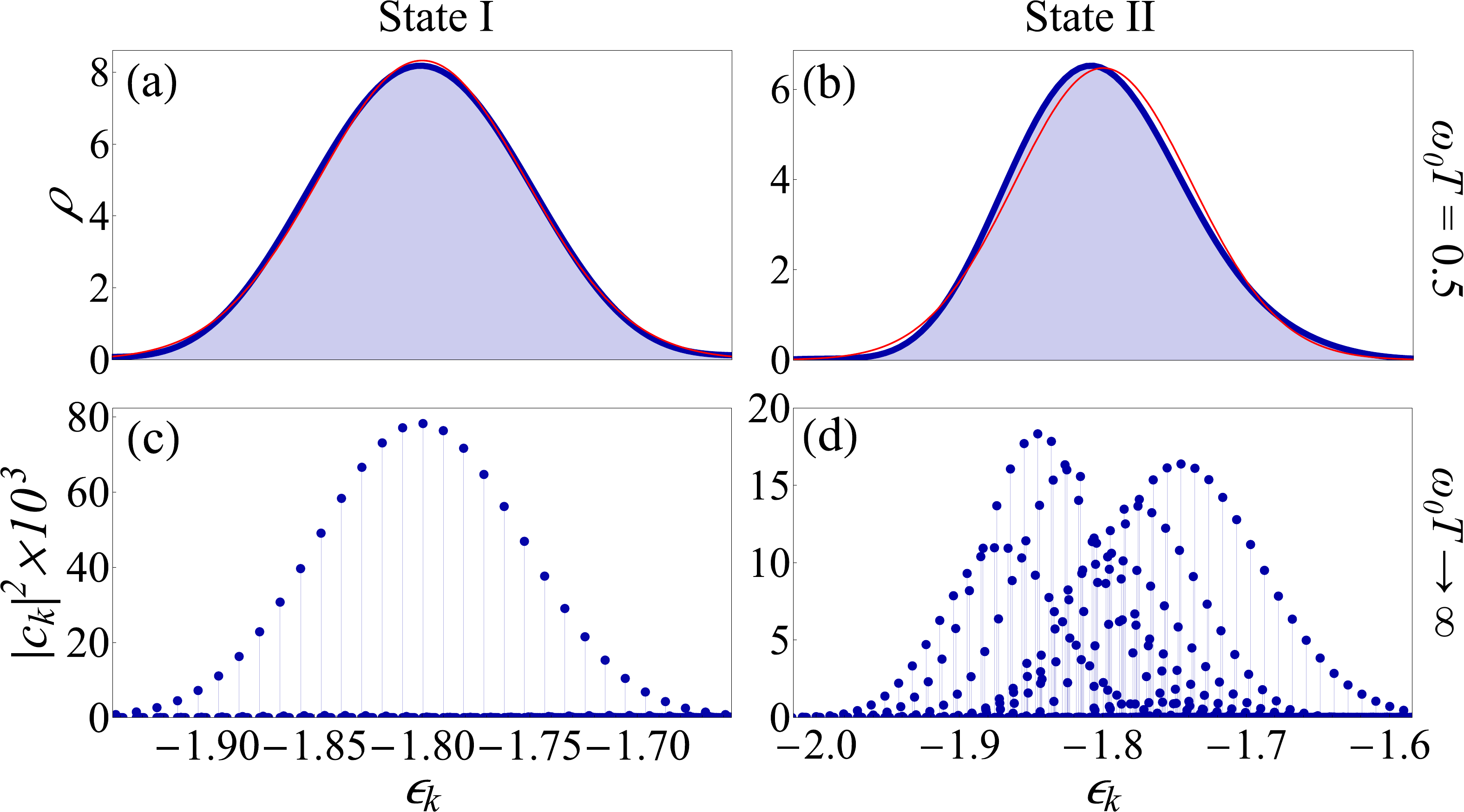}
}
\caption{Smoothed local density of states $\rho_T(E)$ (a,b) and infinite-time resolution LDoS  $\mathcal{G}(E)$ (c,d) for the coherent states I (a,c) and II (b,d) in  the  regular region [see Fig.~\ref{fig:1}~(a)]. In (a) and (b): The time resolution is $\omega_{0}T=0.5$ [$\Delta/(\omega_{0}j)=0.063$]. Solid blue lines are the smoothed energy profiles $\rho_{T}(E)$, given by Eq.~\eqref{eqn:ldos_res}. Red solid lines are the Gaussian energy profiles $\rho(E)$, given by Eq.~\eqref{eqn:gaussian_LDoS}. In (c) and (d): $\omega_{0}T\rightarrow\infty$ [$\Delta/(\omega_{0}j)\rightarrow0$]. Blue dots represent the numerical components of the LDoS. Both profiles are centered at $\epsilon_{R}=-1.8$ with standard deviation $\sigma/(\omega_{0}j)=0.048$ for state I and $\sigma/(\omega_{0}j)=0.062$ for state II. 
}
\label{fig:2}
\end{figure}

\begin{figure}[h!]
\centering{
\includegraphics[width=13cm]{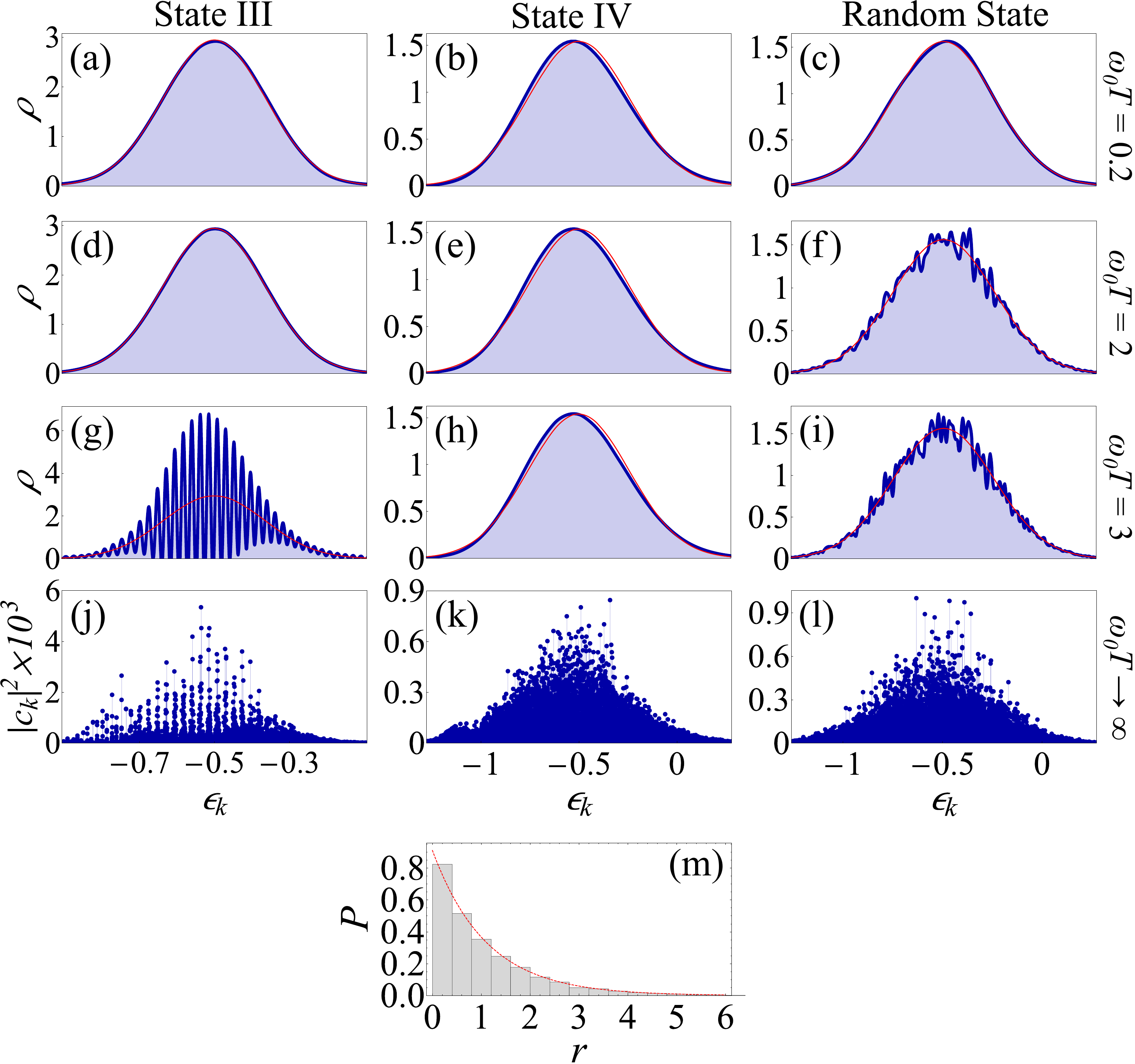}
}
\caption{Smoothed local density of states $\rho_T(E)$ (a)-(i) and infinite-time resolution LDoS $\mathcal{G}(E)$ (j)-(l) for the coherent states III (a,d,g,j) and IV (b,e,h,k) in the chaotic region [see Fig.~\ref{fig:1}~(d)], and a random state (c,f,i,l) (see text). In (a)-(c): The time resolution is $\omega_{0}T=0.2$ [$\Delta/(\omega_{0}j)=0.16$]. In (d)-(f): $\omega_{0}T=2$ [$\Delta/(\omega_{0}j)=0.016$]. In (g)-(i): $\omega_{0}T=3$ [$\Delta/(\omega_{0}j)=0.010$]. Solid blue lines are the smoothed energy profiles $\rho_{T}(E)$ [Eq.~\eqref{eqn:ldos_res}], and red solid lines are the Gaussian energy profiles $\rho(E)$ [Eq.~\eqref{eqn:gaussian_LDoS}]. In (j)-(l): $\omega_{0}T\rightarrow\infty$ [$\Delta/(\omega_{0}j)\rightarrow0$]. Blue dots are the numerical components of the LDoS. All profiles are centered at $\epsilon_{C}=-0.5$ with standard deviation $\sigma/(\omega_{0}j)=0.136$ for state III and $\sigma/(\omega_{0}j)=0.259$ for both the state IV and the random state. Panel (m): Histogram of the numbers $r_k^{\text{CS}}$ (light gray bars) obtained with Eq. \eqref{eqn:cs_rks} for the coherent state IV. The red dashed curve represents the best fit to an exponential distribution $P(r)=\lambda e^{-\lambda r}$, where $\lambda = 0.91$.
}
\label{fig:3}
\end{figure}

Figures~\ref{fig:2}~(a) and (b) and Figs.~\ref{fig:3}~(a), (b), (d), (e), (g), (h) show the smoothed LDoS, $\rho_{T}(E)$, of the initial coherent states described in Sec.~\ref{sec:QuanDyn} for different time resolutions [Eq.~\eqref{eqn:ldos_res}].
In Figs.~\ref{fig:2}~(c) and (d) and Figs.~\ref{fig:3}~(j) and (k), we show the infinite-time resolution LDoS, $\mathcal{G}(E)$.

%%%%%%%%%%%%%%%%%%%%%%%%%%%%%%%%%%%%%%%%%%%%%%%%%%%%%%%
%%%%%%%%%%%%%%%%%%%%%%%%%%%%%%%%%%%%%%%%%%%%%%%%%%%%%%%
%
%%%%%%%%%%%%%%%%%%%%%%%%%%%%%%%%%%%%%%%%%%%%%%%%%%%%%%%
%%%%%%%%%%%%%%%%%%%%%%%%%%%%%%%%%%%%%%%%%%%%%%%%%%%%%%%

%%%%%%%%%%%%%%%%%%%%%%%%%%%%%%%%%%%%%%%%%%%%%%%%%%%%%%%
\subsubsection{Survival Probability: Initial Decay}
\label{subsubsec:Initial_SP_Decay}

For all cases, up to a time resolution $T^{\ast}$ that depends on the state, we find a very good Gaussian distribution for the coherent states profiles,
\begin{equation}
\label{eqn:gaussian_LDoS}
\rho (E)=\frac{1}{\sqrt{2\pi \sigma^2}}e^{-(E-E_{c})^2/(2 \sigma^{2})}.
\end{equation}
The distributions are centered at the energy $E_{c}$ of the initial state and have width given by the energy standard deviation $\sigma$, which can be calculated numerically or even analytically~\cite{Torres2014NJP,Schliemann2015, Lerma2018}.

According to Eq.~(\ref{eq:SPFourier}), the Gaussian envelope of the LDoS leads to an initial Gaussian decay of the survival probability,
\begin{equation}
\label{eqn:gaussian_sp_decay}
S_{P}(t)=e^{-\sigma^{2}t^{2}},
\end{equation}
which is consistent with the universal quadratic behavior, $S_{P} (t\ll\sigma^{-1}) \approx 1-\sigma^{2}t^{2}$, for very short times. 

In addition to the initial coherent states, we show in Figs.~\ref{fig:3}~(c), (f), (i), and (l) the smoothed and infinite-time resolution LDoS of a random initial state. Motivated by the high level of delocalization of the initial coherent state IV in the chaotic regime, the random state is built with energy components generated randomly around the Gaussian distribution of state IV as
\begin{equation}
|c_{k}^{(r)}|^2=\frac{r_{k}\rho(E_{k})}{\mathcal{A}\nu(E_{k})}.
\label{eq:rancom}
\end{equation}
Above, $\rho(E_{k})$ is obtained from Eq.~(\ref{eqn:gaussian_LDoS}), $\nu(E_{k})$ is given by the equation for the density of states (DoS) provided in the \ref{app:1}, $r_k$ are random numbers  from an exponential distribution $P(r)=\lambda e^{-\lambda r}$, and $\mathcal{A}=\sum_{q}r_{q}\rho(E_{q})/\nu(E_{q})$ is a normalization constant. The division by the DoS is done to compensate for the different energy densities and to guarantee a smooth enveloping distribution $\rho(E)$. The exponential distribution for generating the random numbers is used because if we evaluate the numbers
\begin{equation}
\label{eqn:cs_rks}
r^{\text{CS}}_{k}=\frac{\mathcal{A} \nu(E_{k})|c_{k}|^2}{\rho(E_{k})},
\end{equation}
for the components $c_k$  of the coherent state IV in the energy eigenbasis, we obtain the histogram in Fig.~\ref{fig:3}~(m), which is very well fitted with an exponential distribution.

By comparing the smoothed LDoS of the state III in Fig.~\ref{fig:3}~(d), the state IV in Fig.~\ref{fig:3}~(e), and the random state in Fig.~\ref{fig:3}~(f), all of them with time resolution $\omega_{0}T=2$, one notices that the smoothed LDoS of the random state already deviates from a Gaussian, which must affect its short-time dynamics. While the initial coherent states III and IV are expected to remain on the Gaussian decay at this time scale, the random state should already diverge from it, as we indeed confirm in Sec.~\ref{sec:4}. For an even higher time resolution, as $\omega_{0}T = 3$ in Figs.~\ref{fig:3}~(g)-(i), the smoothed LDoS of state III also deviates from the Gaussian, and so does its Gaussian decay (see Sec.~\ref{sec:4}).

%%%%%%%%%%%%%%%%%%%%%%%%%%%%%%%%%%%%%%%%%%%%%%%%%%%%%%%
\subsubsection{Survival Probability: Intermediate Times}

The behavior of the survival probability at intermediate times for the different initial coherent states  can be  anticipated from the infinite-time resolution LDoS in Figs.~\ref{fig:2}~(c) and (d) and Figs.~\ref{fig:3}~(j) and (k). For the regular coherent state I [Fig.~\ref{fig:2}~(c)], very few components have no negligible values and the LDoS is well described by a Gaussian distribution. For the regular coherent state II located close to the separatrix [Fig.~\ref{fig:2}~(d)], the number of participating energy levels is larger, but they are still organized according to a set of different Gaussians with different amplitudes and centers~\cite{Lerma2018}. In contrast, the components of the chaotic coherent states in Fig.~\ref{fig:3} have a very different structure. For the coherent state III, with small $P_R$ [Fig.~\ref{fig:3}~(j)], the components are bunched around some specific energy levels, while the coherent state IV, with a large $P_R$ [Fig.~\ref{fig:3}~(k)], counts with the participation of most components. The consequences of these distributions to the evolution of the survival probability are discussed in Sec.~\ref{sec:4}.

%%%%%%%%%%%%%%%%%%%%%%%%%%%%%%%%%%%%%%%%%%%%%%%%%%%%%%%
\subsubsection{Survival Probability: Asymptotic Values}

The asymptotic value of the survival probability, 
\begin{equation}
\label{eqn:sp_timeaverage}
S_{P}^{\infty}=\langle S_{P}(t)\rangle_{t\rightarrow\infty}=\lim_{t\rightarrow\infty}\frac{1}{t}\int_{0}^{t}\dif t' \, S_{P}(t'),
\end{equation}
can be derived from
\begin{equation}
S_{P}(t)=\sum_{k\neq l}|c_{l}|^{2}|c_{k}|^{2}e^{-i (E_{k}-E_{l})t}+\sum_{k} |c_{k}|^{4}.
\label{Eq:SPrhs}
\end{equation}
In the absence of energy degeneracies, the first term on the right-hand-side of the equation cancels out on average, so 
\begin{equation}
S_{P}^{\infty} = \sum_{k} |c_{k}|^{4},
\end{equation}
which is the inverse of the participation ratio shown in Eq.~(\ref{Eq:PR}).
However, when the  energy levels have degeneracies of degree $d_k$ (that is, $|E_k,m\rangle$ with $m=1,..,d_k$), the first term in Eq.~(\ref{Eq:SPrhs}) contributes with additional terms  to the asymptotic value, which is now given by 
\begin{equation}
S_{P}^\infty=\sum_{E_k}\left(\sum_{m=1}^{d_k}|c_{k,m}|^2\right)^{2},
\label{eqn:spi_no_degenerate}
\end{equation}
where $c_{k,m}$ are the components in the degenerate space of $E_k$.
In the superradiant phase of the Dicke model, for the energy range going from the ground state to $\epsilon=-1$, the spontaneous breaking of the parity symmetry~ \cite{Puebla2013} produces an energy spectrum with two-fold degeneracies\footnote{Because of tunneling effects in the energy region with $\epsilon<-1$, parity partners are not exactly degenerated, but we verified that their energy differences are smaller than the numerical precision of our numerical calculations.}.  For the initial coherent states in the regular region, these degeneracies affect their asymptotic values, while for the chaotic initial states, where the energy $\epsilon_C$ is well above $\epsilon=-1$, the influence of the degeneracies is rather marginal.

%%%%%%%%%%%%%%%%%%%%%%%%%%%%%%%%%%%%%%%%%%%%%%%%%%%%%%%
%%%%%%%%%%%%%%%%%%%%%%%%%%%%%%%%%%%%%%%%%%%%%%%%%%%%%%%
\subsection{Classical Limit of the Survival Probability}

To find the classical limit of the survival probability, we use the Wigner formalism \cite{Wigner1932}. Details are given in \ref{app:3} and \ref{app:4}. The basic idea is to use the overlap property between two arbitrary quantum states $\ket{A}$ and $\ket{B}$ to write the survival probability in terms of the Wigner function $W$ \cite{Case2008}, 
\begin{equation}
\abs{\braket{A}{B}}^2=\left (2\pi\hbar \right)^d \int{\dif  \bm u \, W_A(\bm u)W_B(\bm u) },
\end{equation}
where $d$ represents the degrees of freedom of the system.
We scale the Wigner function of our Glauber-Bloch initial coherent states and work with $w(\bm u,t)=(1/j)W(\sqrt{j}q,\sqrt{j}p,Q,P,t)$, where $\boldsymbol u=(q,p,Q,P)$ is a point in the $j$-scaled phase space $\mathcal{M}$. This gives
\begin{equation}
\label{eqn:SPWigner}
S_P(t)=\left(\frac{2\pi}{j}\right)^2\int_{\mathcal{M}} \dif \bm{u}\, w(\bm{u},0) w(\bm{u},t).
\end{equation}
To finally obtain the classical limit, we use the TWA. As shown in \ref{app:4}, the short-time dependence of the Wigner function can be written in terms of the Hamiltonian flow $\flow{t}:\mathcal{M} \to \mathcal{M}$. One finds that $w(\bm{u},t)=w(\flow{-t}(\bm{u}),0)$ for short times, so the classical survival probability can be defined as
\begin{equation}
\label{eqn:SPconflujo}
\mathfrak{S}_{P}(t)= \left(\frac{2\pi}{j} \right) ^2\int_{\mathcal{M}} \dif \bm{u}\,  w(\bm{u}) w(\flow{-t}(\bm{u})),
\end{equation}
where $w(\bm{u})=w(\bm{u},t=0)$. This quantity is numerically constructed through a Monte Carlo method (see \ref{app:4}). Its asymptotic value, $\mathfrak{S}_{P}^\infty = \expval{\mathfrak{S}_{P}(t)}_{t\to\infty}$,  is obtained as in Eq.~\eqref{eqn:sp_timeaverage}.

%%%%%%%%%%%%%%%%%%%%%%%%%%%%%%%%%%%%%%%%%%%%%%%%%%%%%%%
%%%%%%%%%%%%%% SEC.4  CLASSICAL AND QUANTUM DYNAMICS %%%%%%%%%%%%%%%%
%%%%%%%%%%%%%%%%%%%%%%%%%%%%%%%%%%%%%%%%%%%%%%%%%%%%%%%
\section{CLASSICAL AND QUANTUM DYNAMICS}
\label{sec:4}

We compare the entire evolution of the quantum [Eq.~\eqref{eqn:qua_sp}] and classical [Eq.~\eqref{eqn:SPconflujo}] survival probabilities for the four initial coherent states selected in Sec.~\ref{sec:QuanDyn}. This analysis allows us to identify features of the quantum evolution that reflects the classical dynamics and properties that are purely quantum. We find that:

--  The $S_P(t)$ for coherent initial states reaches values close to zero at short times. This behavior can be understood from the short-time evolution of the Wigner distribution in phase space, and is thus a classical effect.

--  The long-time behavior of initial states with the same energy in the regular or chaotic region may differ according to the states' level of delocalization in the energy eigenbasis.

-- Purely quantum properties of the survival probability include not only the quantum fluctuations of $S_P(t)$ after the saturation of the dynamics, but also tunneling effects, the dynamical consequences of the phenomenon of quantum scarring, and the dynamical manifestation of spectral correlations in the form of the correlation hole.

-- The saturation times of the quantum and classical dynamics in the chaotic regime do not coincide. Quantum equilibration is shorter, when the quantum initial state has large projections in scarred eigenstates, reflecting the proximity to unstable periodic orbits in phase space. Quantum equilibration is longer, when the initial state is highly delocalized and $S_P(t)$ develops the correlation hole before saturation.

-- The value of the ratio between the asymptotic values of the quantum and classical survival probabilities indicates whether the initial state is close to a separatrix and whether it is close to an unstable periodic orbit. A ratio equal to two indicates maximal quantum ergodicity.

%%%%%%%%%%%%%%%%%%%%%%%%%%%%%%%%%%%%%%%%%%%%%%%%%%%%%%%
%%%%%%%%%%%%%%%%%%%%%%%%%%%%%%%%%%%%%%%%%%%%%%%%%%%%%%%
\subsection{Regular Region}

We analyze first the quantum $S_P(t)$ and the classical $\mathfrak{S}_{P}(t)$ for the initial coherent states I and II [Fig.~\ref{fig:1}~(a)], which are in the regular regime ($\epsilon_R=-1.8$).

%%%%%%%%%%%%%%%%%%%%%%%%%%%%%%%%%%%%%%%%%%%%%%%%%%%%%%%
\subsubsection{Initial State I: Center of the Regular Mode}

An analytical equation is available for the survival probability of the initial coherent state I,  located at the center of the slow-boson regular region~\cite{Lerma2018}. It is given by
\begin{equation}
S_P(t)  \approx 
\frac{\omega_1}{\sigma\sqrt{\pi}} \displaystyle\sum_{n=1} 
 e^{ -n^2\left( \frac{ \omega_1^2}{4 \sigma^2} +  \frac{t^2}{t_D^2} \right)}  \cos (n \omega_1 t) + S_P^{\infty},
\label{eqn:sp_regular}
\end{equation}
where the index $n$ denotes the distance between the eigenenergies of the levels with no negligible components, $\omega_1$ is related with the spacing between neighboring levels and is equal to the average of the classical frequency  over the initial Wigner distribution, $\sigma$ is the width of the LDoS, $t_{D}=\omega_{1}/(\sigma |e_{2}|)$ is the decay time, with $e_2$ being the anharmonicity of the participating spectrum, and 
\begin{equation}
S_P^{\infty}=\frac{\omega_1}{2\sigma\sqrt{\pi}}.
\label{eq:IPRA}
\end{equation}
The time $t_{D}$ comes from the Gaussian decay of the slowest component $n=1$.

%%%%%%%%%%%%%%%%%%%%%%%%%%%%%%%%%%%%%%%%%%%%%%%%%%%%%%%
\begin{figure}[h!]
\centering{
\includegraphics[width=12cm]{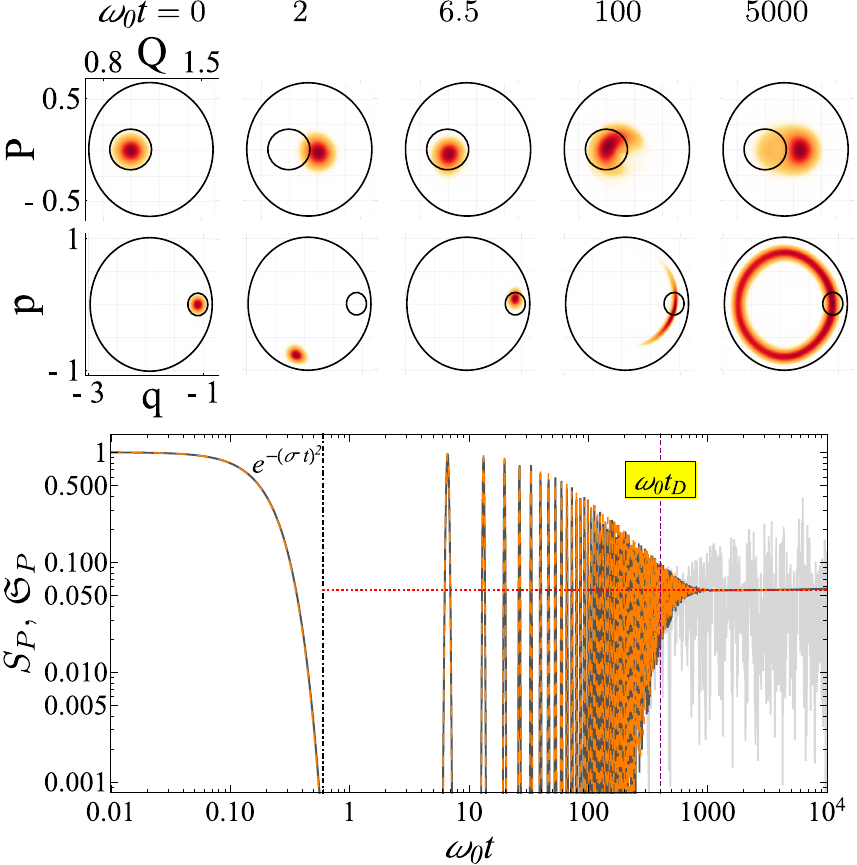}
}
\caption{Top panels: Classical evolution of the Wigner distribution projected onto the $Q$-$P$ plane (top row) and $q$-$p$ plane (bottom row) for the initial coherent state I in Fig.~\ref{fig:1}~(a). The initial distribution (small black circle) is inside the available phase space (big black circle). Each column represents an instant of time, as indicated. (Animations can be found in the SM~\cite{SM}.)
Bottom large panel: Quantum survival probability (light gray solid line), analytical expression from Eq.~\eqref{eqn:sp_regular} (orange dashed line), and classical survival probability (dark gray solid line) for the initial coherent state I. The horizontal red dotted line shows the asymptotic value $S_P^{\infty}$ [Eq.~\eqref{eq:IPRA}]. The decay time $t_{D}$ is indicated with a vertical line. The analytical expression uses the parameters from the numerical data $(\sigma,\omega_{1},e_{2})=(4.79,0.94,-4.88\times10^{-4})$, which gives $\omega_{0}t_{D}=403.8$.
}
\label{fig:4}
\end{figure}
%%%%%%%%%%%%%%%%%%%%%%%%%%%%%%%%%%%%%%%%%%%%%%%%%%%%%%%

In the bottom panel of Fig.~\ref{fig:4}, we compare the full quantum evolution of $S_P(t)$ with the analytical expression in Eq.~(\ref{eqn:sp_regular}) and the classical $\mathfrak{S}_{P}(t)$. The latter two show perfect agreement and they are indistinguishable from the quantum result up to the decay time $t_{D}$. Soon after this point, $S_P(t)$ shows quantum fluctuations around $S_P^{\infty}$ that emerge due to the discreteness of the spectrum and which the analytical expression in Eq.~\eqref{eqn:sp_regular} and $\mathfrak{S}_{P}(t)$ are unable to reproduce.

The remarkable agreement between the quantum and classical results allows for a more intuitive classical interpretation of the behavior of the survival probability. In the top panels of Fig.~\ref{fig:4}, we show the classical evolution of the Wigner distribution in the phase space. The top row gives the distribution projected onto the $Q$-$P$ plane and the bottom row, the distribution on the $q$-$p$ plane. Each column corresponds to a specific instant of time, as indicated. The classical survival probability measures the percentage of the Wigner distribution (in color in the Figure) that is inside of the starting region occupied at $t=0$ (See \ref{app:6} for a detailed explanation). This starting region is marked with a black outline which is inside a bigger one marking the available phase space. The parallel between the evolution of the survival probability and of the Wigner distribution in the phase space goes as follows (animations are available in the Supplemental Material (SM)~\cite{SM}):

-- At $\omega_0 t=0$ (first column), the distribution is entirely within the starting region. Concurrently, the value of the survival probability in the bottom panel is one. 

-- As the distribution moves out of the starting region, the survival probability follows a Gaussian decay. At $\omega_0 t=2$ (second column), the distribution is effectively outside the initial region and the survival probability becomes close to zero\footnote{Note that even though in the $Q$-$P$ projection the distribution appears to be still partially inside the starting region, this is merely an artifact caused by the projection. In the $q$-$p$ projection, the distribution is already clearly outside the starting region. This should be kept in mind when interpreting these figures, both projections are important and cannot be treated independently.}. 
 
-- After a full period given by $t=2\pi/\omega_1$, the distribution comes back to the starting region at $\omega_0 t=6.5$ (third column) and we have the first revival of $S_P(t)$. 

-- After several periods, the distribution further spreads over the classical orbit and the amplitude of the revivals of the survival probability decreases, as for example at $\omega_0 t=100$ (fourth column).

-- Classically, at $\omega_0 t=5000$ (fifth column), the Wigner distribution ends up filling homogeneously the region covered by the trajectories of the phase-space points of the initial distribution. At this point the classical $\mathfrak{S}_{P}(t)$ reaches the  constant value $\mathfrak{S}_{P}^\infty$, while $S_P(t)$ fluctuates around the asymptotic value $S_{P}^{\infty}\approx \mathfrak{S}_{P}^{\infty}$.

%%%%%%%%%%%%%%%%%%%%%%%%%%%%%%%%%%%%%%%%%%%%%%%%%%%%%%%
\subsubsection{Initial State II: Close to the Separatrix}

The LDoS of the initial coherent state II presents different Gaussian distributions with different amplitudes, centers and widths [Fig.~\ref{fig:2}~(d)]. This indicates that this state activates the two adiabatic modes (both the slow boson mode and the fast pseudo-spin mode), as well as the non-linear resonances between them. This is indeed expected, since state II is very close to the separatrix [Fig.~\ref{fig:1}~(a)]. An analytical expression can also be obtained for this case by generalizing Eq.~\eqref{eqn:sp_regular}, where in addition to contributions from Gaussians, we also need to take into account interferences between them~\cite{Lerma2018}.
 
\begin{figure}[h!]
\centering{
\includegraphics[width=12cm]{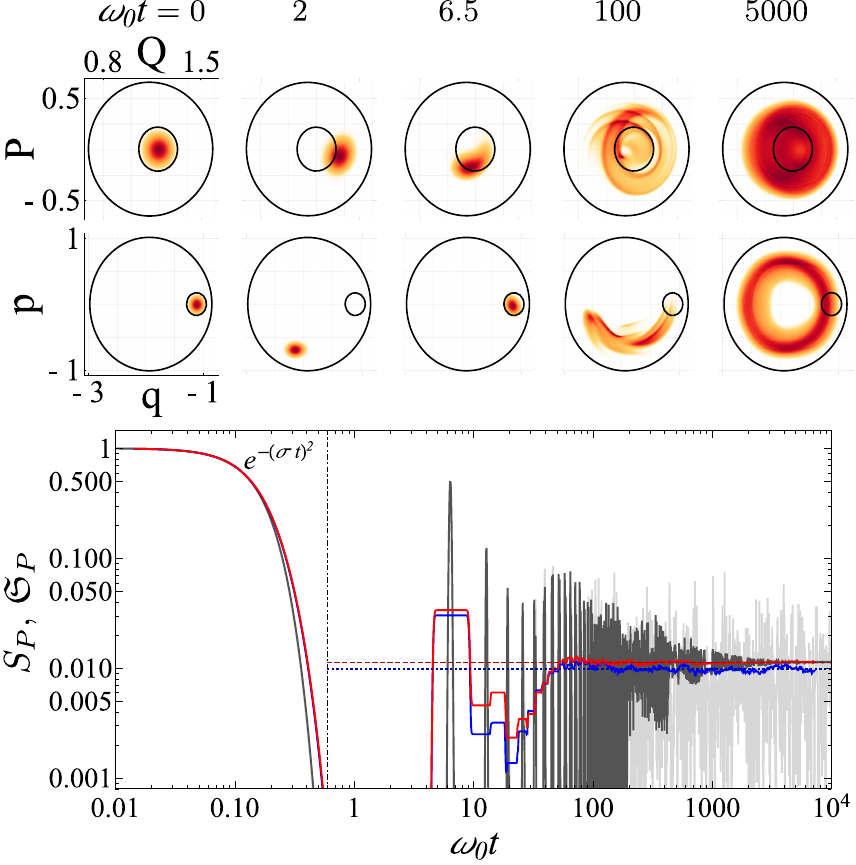}
}
\caption{Top panels: Classical evolution of the Wigner distribution projected onto the $Q$-$P$ plane (top row) and $q$-$p$ plane (bottom row) for the initial coherent state II from Fig.~\ref{fig:1}~(a). The initial distribution (small black circle) is inside the available phase space (big black circle). Each column represents a time, as indicated. (Animations can be found in the SM~\cite{SM}.)
Bottom panel: Quantum survival probability (light gray solid line), its temporal average (blue solid line), classical survival probability (dark gray solid line), and its temporal average (red solid line) for the initial coherent state II. The temporal averages are computed for temporal windows of constant size in the logarithmic scale. The horizontal blue dotted line is the quantum asymptotic value $S_{P}^{\infty}$ and the red dashed line corresponds to the classical asymptotic value $\mathfrak{S}_{P}^\infty$.
}
\label{fig:5}
\end{figure}

In the bottom panel of Fig.~\ref{fig:5}, we show the quantum and classical survival probability for the state II. The agreement between the two is good, but  contrary to Fig.~\ref{fig:4}, the values of $S_P(t)$ are slightly smaller than those for $\mathfrak{S}_{P}(t)$. This is better seen with the temporal averages of the curves, shown in Fig.~\ref{fig:5} in blue for $S_P(t)$ and in red for $\mathfrak{S}_{P}(t)$.

%%%%%%%%%%%%%%%%%%%%%%%%%%%%%%%%%%%%%%%%%%%%%%%%%%%%%%%
%%%%%%%%%%%%%%%%%%%%%%%%%%%%%%%%%%%%%%%%%%%%%%%%%%%%%%%
%
%%%%%%%%%%%%%%%%%%%%%%%%%%%%%%%%%%%%%%%%%%%%%%%%%%%%%%%
\begin{figure}[h!]
\centering{
\includegraphics[width=10cm]{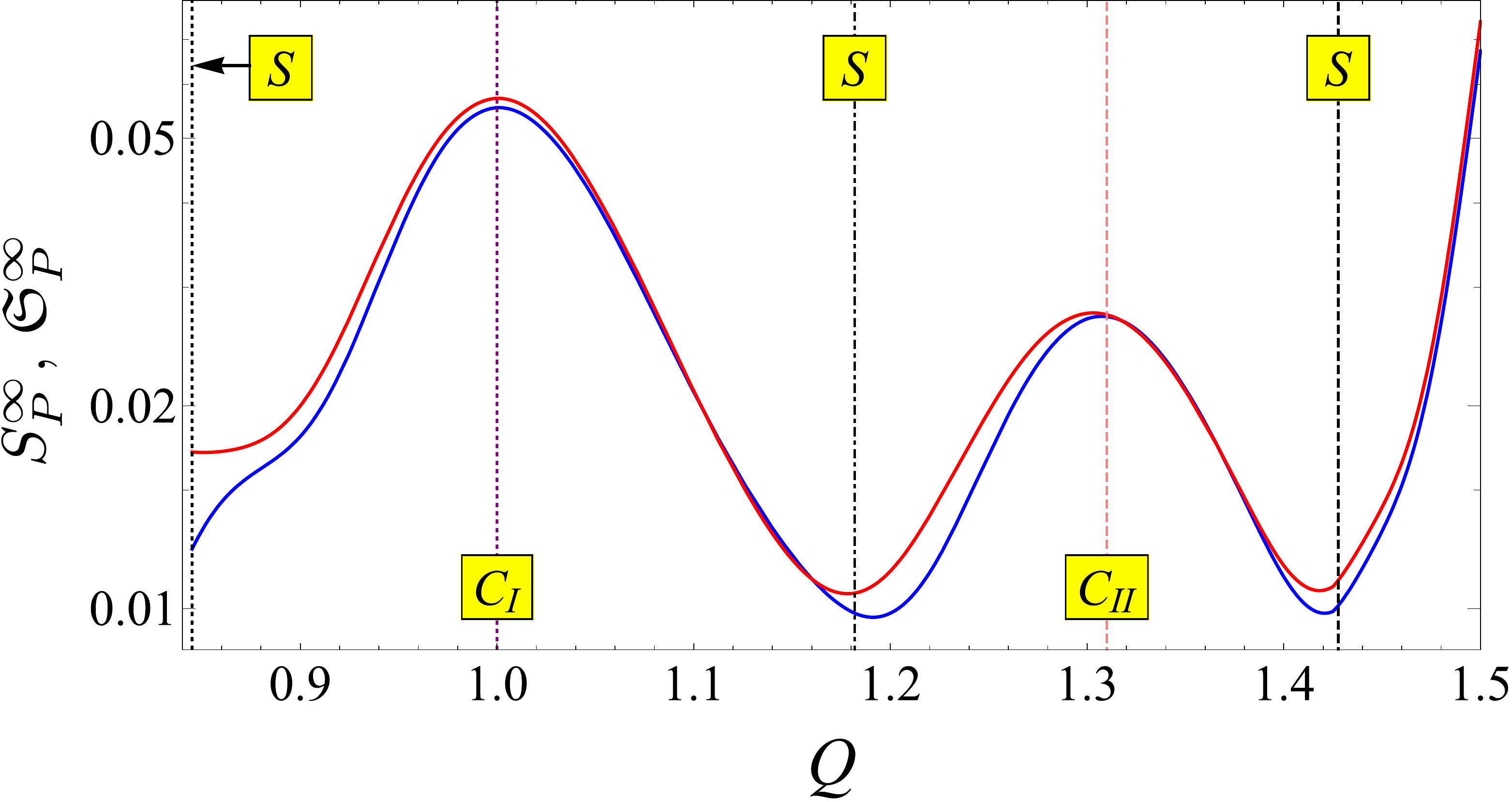}
}
\vskip -0.3 cm
\caption{Quantum asymptotic value $S_{P}^{\infty}$ (blue solid line) and classical asymptotic value $\mathfrak{S}_{P}^{\infty}$ (red solid line) for a set of initial coherent states in the same regular Poincar\'e section of Fig.~\ref{fig:1}~(a) with $P=0$ and $Q$ covering the whole allowed interval. The vertical lines marked with $S$ indicate the separatrix at different crossing points, the one marked with  $C_{I}$ indicates the center of an adiabatic-mode region and the one marked with  $C_{II}$ indicates the center of the non-linear resonances.
}
\label{fig:6}
\end{figure}
%%%%%%%%%%%%%%%%%%%%%%%%%%%%%%%%%%%%%%%%%%%%%%%%%%%%%%%

To better understand the small differences between $S_P(t)$ and $\mathfrak{S}_{P}(t)$, we  calculate their asymptotic values for initial coherent states $(q,p,Q,P)=(q_{+},0,Q_{i},0)$, where $Q_{i}$ covers all possible points of the selected Poincar\'e surface in Fig.~\ref{fig:1}~(a). We plot the values of $S_P^{\infty}$ and  $\mathfrak{S}_{P}^{\infty}$ in Fig.~\ref{fig:6}. 
One sees that $S_{P}^{\infty}$ is smaller than $\mathfrak{S}_{P}^{\infty}$ near the separatrix regions ($Q=0.845,1.182$, and $1.428$), while the two values get closer near the centers of the adiabatic-modes regions ($Q=1.0, 1.5$) and of the center of the non-linear resonances ($Q=1.31$). Since a coherent state located in the separatrix has a Wigner distribution defined in regions classically not connected by the trajectories, we attribute the small difference between $S_P(t)$ and $\mathfrak{S}_{P}(t)$  to a dynamic tunneling effect that takes place in the quantum regime~\cite{KeshavamurthyBook}, but is absent in the classical limit. 

Even though the quantum-classical agreement for state II is not exact, the comparison between the results for the survival probability in the bottom panel and the evolution of the Wigner distribution in the top panels is similar to that presented in Fig.~\ref{fig:4} and explains specific features of $S_P(t)$, such as the revival at $\omega_0 t= 6.5$. At very long times, such as $\omega_0 t=5000$ in the figure, the Wigner distribution fills homogeneously the classical region that is covered by the trajectories of the initial distribution. Since this region is larger than in the case of state I, the asymptotic value in Fig.~\ref{fig:5} is one order of magnitude smaller than in Fig.~\ref{fig:4}.

%%%%%%%%%%%%%%%%%%%%%%%%%%%%%%%%%%%%%%%%%%%%%%%%%%%%%%%
%%%%%%%%%%%%%%%%%%%%%%%%%%%%%%%%%%%%%%%%%%%%%%%%%%%%%%%
\subsection{Chaotic Region}

We now study the quantum $S_P(t)$ and the classical $\mathfrak{S}_{P}(t)$ for the initial coherent states III and IV [Fig.~\ref{fig:1}~(d)], which are in the chaotic regime ($\epsilon_C=-0.5$). We start the analysis with state III, which has a smaller number of contributing energy eigenstates (smaller $P_R$) than state IV.

%%%%%%%%%%%%%%%%%%%%%%%%%%%%%%%%%%%%%%%%%%%%%%%%%%%%%%%
\subsubsection{Initial State III: Low $P_R$}

%%%%%%%%%%%%%%%%%%%%%%%%%%%%%%%%%%%%%%%%%%%%%%%%%%%%%%%
\begin{figure}[h!]
\centering{
\includegraphics[width=12cm]{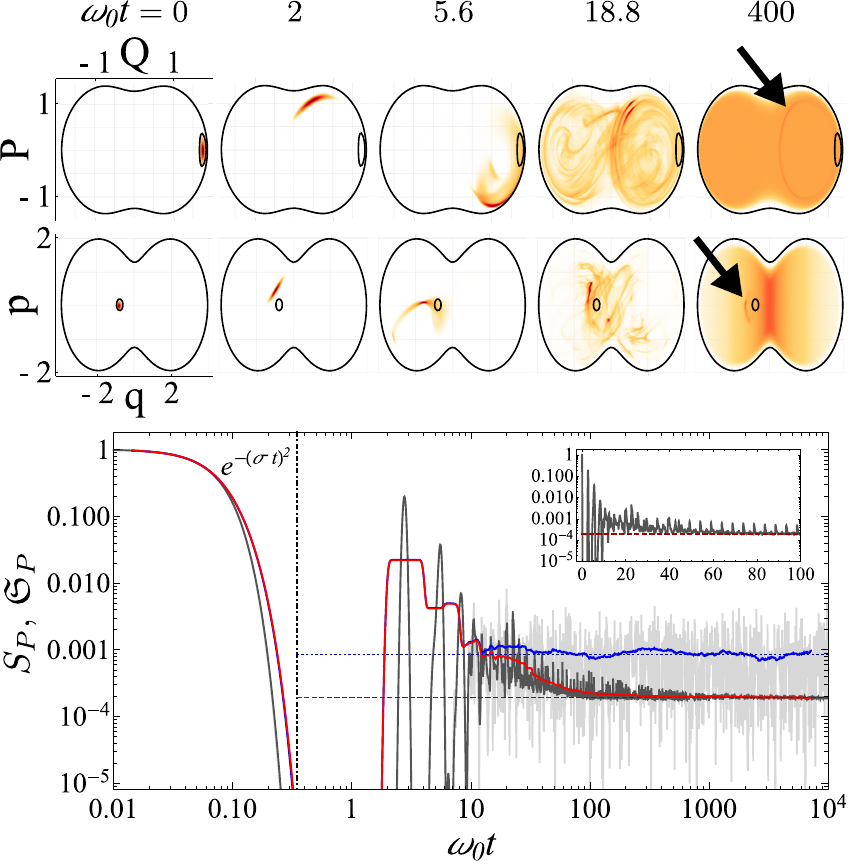}
}
\caption{Top panels: Classical evolution of the Wigner distribution projected onto the $Q$-$P$ plane (top row) and $q$-$p$ plane (bottom row) for the initial coherent state III from Fig.~\ref{fig:1}~(d). The initial distribution (small black circle) is inside the available phase space. Each column represents a time, as indicated. (Animations can be found in the SM~\cite{SM}.) The black arrows indicate an enhancement of the distribution around an unstable periodic orbit (see text).
Bottom panel: Quantum survival probability (light gray solid line), its temporal average (blue solid line), classical survival probability (dark gray solid line), and its temporal average (red solid line) for the initial coherent state III. The temporal averages are computed for temporal windows of constant size in the logarithmic scale. The horizontal blue dotted line is the quantum asymptotic value $S_{P}^{\infty}$ and the horizontal red dashed line corresponds to the classical asymptotic value $\mathfrak{S}_{P}^\infty$.
The inset shows the classical approximation (TWA) of the survival probability in a semi-log scale, which makes clear that the oscillations are periodic. 
}  
\label{fig:7}
\end{figure}
%%%%%%%%%%%%%%%%%%%%%%%%%%%%%%%%%%%%%%%%%%%%%%%%%%%%%%%

Despite being in the chaotic region, initial coherent states such as state III count with a relatively small number of contributing energy eigenstates, as indicated by the low value of $P_R$ in Fig.~\ref{fig:1}~(d). As we show below, these states lead to large recurrences at intermediate times and, contrary to the regular case, the equilibration time for the classical and quantum dynamics no longer coincide. The equilibration time for the classical dynamics is now longer than the equilibration time for the quantum $S_P(t)$.

The quantum and classical survival probabilities of the initial coherent state III are shown in the bottom panel of Fig.~\ref{fig:7}. There is excellent agreement from $\omega_{0}t=0$ up to times beyond the decaying recurrences. Initially both curves decay on a Gaussian to values close to zero and then revivals appear. These two features can be well understood by studying the classical evolution of the Wigner distribution shown in the top panels of Fig.~\ref{fig:7}. Analogously to the discussions about Fig.~\ref{fig:4} and Fig.~\ref{fig:5}, as the originally localized Wigner distribution at $\omega_0 t=0$ (first column) moves outside the starting region at $\omega_0 t=2$ (second column), the survival probability becomes effectively zero. The recurrences are connected with the return of the distribution to the starting region, such as at $\omega_0 t=5.6.$ (third column). These recurrences are periodic, as confirmed with the inset in the bottom panel of Fig.~\ref{fig:7}.

Soon after the last revival, the quantum-classical correspondence breaks down. At this point the quantum $S_P(t)$ reaches an asymptotic value, around which one finds large quantum fluctuations, while the classical $\mathfrak{S}_{P}(t)$ continues decreasing and equilibrates at a longer time. The asymptotic value of the classical survival probability is obtained using the TWA and the ergodic hypothesis (see \ref{app:5}),
\begin{equation}
\label{eqn:resSPCAsymptapprox}
\mathfrak{S}_{P}^\infty=\frac{1}{2 \sqrt{\pi} \sigma \nu_{c}}.
\end{equation}
The asymptotic result $S_P^{\infty}$ of the quantum survival probability is more than twice this value.

Large $S_P^{\infty}$, or equivalently small $P_R$, in the chaotic region is usually associated with the phenomenon of scarring, which is indeed the case here. The classical Wigner distribution at $\omega_0 t=400$ (fifth column in the top panel of Fig.~\ref{fig:7}) is enhanced in a small closed region indicated with black arrows in the figure. This reveals the presence of  unstable periodic orbits of relatively short period. These orbits are responsible for the short-time periodic revivals of the quantum $S_P(t)$ and classical $\mathfrak{S}_{P}(t)$. 

For longer times, the periodic orbits produce opposing effects in the quantum and classical regimes.
Classically, we observe a slow decay of $\mathfrak{S}_{P}(t)$ towards its asymptotic value and thus a long classical equilibration time if compared to the quantum case. Recurrences imply that the dynamics revisits part of the phase space that was initially covered instead of exploring new regions, which slows down the full spread over the phase space~\cite{Heller1987,Heller1991}. 
In the quantum evolution, the scarring decreases the equilibration time. As discussed in Ref.~\cite{Heller1984} and as can be seen in Fig.~\ref{fig:3}~(g), the proximity of an initial state to unstable periodic orbits of short period  produces a particular structure in the distribution of the energy eigenbasis components. Many of these components are very small, while the large ones are organized in bunches separated in energy by $\Delta E_s\approx 2 \pi/\tau$, where $\tau$ is the period of the classical unstable periodic orbit. 
These highly populated eigenstates are \textit{scarred}, meaning that they are concentrated in the phase space around the unstable periodic orbits~\cite{Heller1987,Heller1991}.  Because of this concentration, the phase space available for the quantum evolution of the initial coherent state is effectively shrunk, resulting in an smaller equilibration time for $S_P(t)$ than for $\mathfrak{S}_{P}(t)$ and an asymptotic quantum value larger than the classical one, $S_{P}^\infty> 2\mathfrak{S}_{P}^\infty$.

%%%%%%%%%%%%%%%%%%%%%%%%%%%%%%%%%%%%%%%%%%%%%%%%%%%%%%%
\subsubsection{Initial State IV: High $P_R$ }

The behavior of the survival probability presented in the previous subsection is not general. Most initial coherent states in the chaotic region are similar to state IV, being highly delocalized in the energy eigenbasis. In fact, as discussed in Sec. \ref{subsect:TimeScales}, the energy distribution of state IV is comparable to that of a random state. This latter state is very useful, because one can derive an analytical expression for its survival probability. The expression for the average over an ensemble of initial random states is given by (see \cite{Lerma2019} and \ref{app:2} for details)
\begin{equation}
\label{eqn:SPRMT}
S_{P}^{(r)}(t)=\frac{1-S_{P}^{(r),\infty}}{\eta-1}\left[\eta e^{-\sigma^{2}t^{2}}-b_{2}\left(\frac{Dt}{2\pi}\right)\right]+S_{P}^{(r),\infty},
\end{equation}
where $\eta=2\sqrt{\pi}\sigma\nu_{c}$ and $\nu_{c}=\nu(E_{c})$ is the DoS [see Eq.~\eqref{eqn:DoS} in  \ref{app:1}] evaluated in the center of the energy profile $E_{c}$, $\sigma$ is the width of the LDoS, the $b_2$ function is the Gaussian orthogonal ensemble (GOE) two-level form factor studied in random matrix theory~\cite{MehtaBook},  $D=2/\nu_c$ is the mean level spacing of the correlated eigenvalues, and $S_P^{(r),\infty}=2/\eta$.  
By comparing the quantum survival probability with $S_P^{(r)}(t)$ and the classical $\mathfrak{S}_{P}(t)$, we can explain the different behaviors of $S_P(t)$ found at different time scales.

\begin{figure}[h!]
\centering{
\includegraphics[width=12cm]{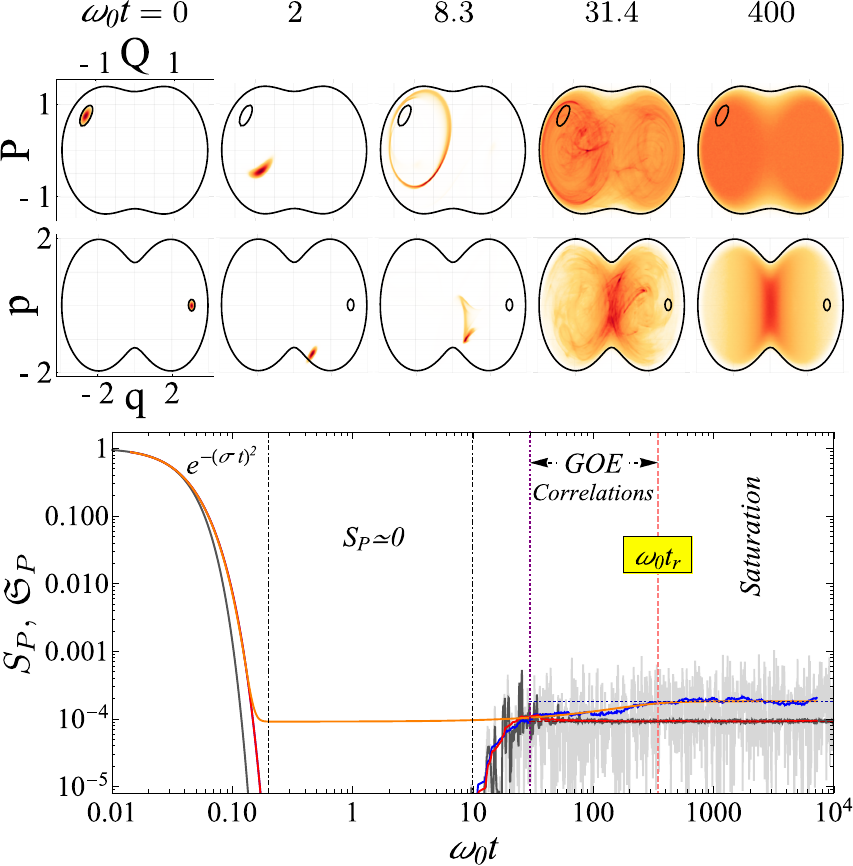}
}
\caption{Top panels: Classical evolution of the Wigner distribution projected onto the $Q$-$P$ plane (top row) and $q$-$p$ plane (bottom row) for the initial coherent state IV from Fig.~\ref{fig:1}~(d). The initial distribution (small black circle) is inside the available phase space. Each column represents a time, as indicated. (Animations can be found in the SM~\cite{SM}.)
Bottom panel: Quantum survival probability (light gray solid line), its temporal average (blue solid line), classical survival probability (dark gray solid line), and its temporal average (red solid line) for the initial coherent state IV, as well as the temporal average of the analytical expression for random states given in Eq.~\eqref{eq:ranas} (orange solid line). The temporal averages are computed for temporal windows of constant size in the logarithmic scale. The horizontal blue dotted line is the quantum asymptotic value $S_{P}^{\infty}$ [Eq.~\eqref{eq:ranas}] and the horizontal red dashed line corresponds to the classical asymptotic value $\mathfrak{S}_{P}^\infty$ [Eq.~\eqref{eqn:resSPCAsymptapprox}]. The vertical line at $\omega_{0}t \approx 30$ marks the correlation hole and beginning of the ramp of $S_P(t)$ towards saturation. The relaxation time $\omega_0 t_{r}=345.7$ [Eq.~\eqref{eq:eqtime}], marked with a vertical line, is calculated using $\delta_{S_{P}}=0.05$.
}
\label{fig:8}
\end{figure}

In the bottom panel of Fig.~\ref{fig:8}, we show the quantum $S_P(t)$ of state IV (light grey solid line), its time average (blue solid line), the classical $\mathfrak{S}_{P}(t)$ (dark grey solid line), its time average (red solid line), and the time average of the analytical expression $S_P^{(r)}(t)$ (orange solid line). At short times, $S_P(t)$ and $\mathfrak{S}_{P}(t)$ overlap. There is also perfect agreement between the Gaussian decay of the temporal average of $\mathfrak{S}_{P}(t)$, $S_P(t)$ and $S_P^{(r)}(t)$, as expected from the great similarity between the smoothed LDoS at low time resolution ($\omega_{0}T=0.2$) of the coherent and the random state [Fig.~\ref{fig:3}~(b) and (c)]. However, the curves for $S_P(t)$ and $S_P^{(r)}(t)$ diverge after $\omega_{0}t\sim 0.2$. One sees that $S_P(t)$ remains on the Gaussian decay, while $S_P^{(r)}(t)$ reaches a plateau. This divergence can also be understood from the smoothed LDoS, but now at higher time resolution ($\omega_{0}T=2$ and $3$) [see Figs.~\ref{fig:3}~(e)(f) and  Figs.~\ref{fig:3}~(h)(i)]: while the smoothed LDoS for the state IV is still a good Gaussian, that of the random state shows deviations. This discrepancy implies that the components of the coherent state IV are not exactly random.

%%%%%%%%%%%%%%%%%%%%%%%%%%%%%%%%%%%%%%%%%%%%%%%%%%%%%%%
%%%%%%%%%%%%%%%%%%%%%%%%%%%%%%%%%%%%%%%%%%%%%%%%%%%%%%%

The quantum and classical survival probabilities continue their  Gaussian decay for $\omega_0 t> 0.2$ and reach values close to zero, which can once again be understood from the analysis of the classical evolution of the Wigner distribution shown in the top panels of Fig.~\ref{fig:8}. As the initially localized Wigner distribution at $\omega_0 t=0$ (first column) moves out of the starting region at $\omega_0 t=2$ and $\omega_0 t=8.3$ (second and third columns), the survival probability becomes effectively zero. Beyond these times and contrary to what one sees for the regular regime and for state III, the quantum-classical correspondence breaks down before the quantum equilibration, as we explain next.

At $\omega_0 t\approx 30$ (marked with a vertical line in the large bottom panel of Fig.~\ref{fig:8}), the classical survival probability attains its asymptotic equilibration value $\mathfrak{S}_{P}^\infty$ and the  Wigner distribution covers the entire energy shell, as seen in the top panels of Fig.~\ref{fig:8} for $\omega_0 t=31.4$ and $\omega_0 t=400$ (fourth and fifth columns). In contrast, the quantum $S_P(t)$ continues to raise, leading to a longer equilibration time. For $\omega_0 t > 30 $, the temporal averaged $S_P(t)$ coincides again extremely well with the analytical expression for $S_P^{(r)}(t)$. During this ramp toward equilibration, the dynamics is controlled by the $b_2$ function, which reflects the correlations between the energy levels in the chaotic regime. For the chaotic Dicke model, these correlations  are equivalent to those in the spectrum of GOE random matrices~\cite{Emary2003PRL,Emary2003,Bastarrachea2014a,Buijsman2017}, which justifies the use in Eq.~(\ref{eqn:SPRMT}) of  the same form of the $b_2$ function from random matrix theory.  At this time scale the dynamics becomes universal. The region of the ramp is referred to in Fig.~\ref{fig:8} as ``GOE correlations'' and is commonly known as correlation hole~\cite{Leviandier1986,Wilkie1991,Alhassid1992,Alt1997,Gorin2002,Torres2017PTR,Torres2018,Schiulaz2019,Lerma2019}. This is a quantum effect associated with the discreteness of the spectrum, which does not appear for the classical $\mathfrak{S}_{P}(t)$.

In contrast to the scarred state III, the quantum survival probability of the state IV reaches its asymptotic value at a time $t_r$ that is  longer than the classical relaxation time, and is given by the same expression obtained with the analytical equation for $S_P^{(r)}(t)$ (see Ref.~\cite{Lerma2019} and \ref{app:2}),
\begin{equation}
t_r=\frac{\pi\nu_c}{2\sqrt{6 \, \delta_{S_P}}},
\label{eq:eqtime}
\end{equation}
where $\delta_{S_P}$ is a small parameter indicating that the values of $S_P(t>t_r)$ are already within the quantum fluctuations around $S_{P}^{\infty}$. This time is marked with a vertical line in Fig.~\ref{fig:8}.

The asymptotic value of the quantum survival probability agrees with the saturation point of $S_P^{(r)}(t)$ (see \ref{app:2}),
\begin{equation}
S_{P}^{\infty}=S_{P}^{(r),\infty}=\frac{\langle r^2\rangle}{\langle r\rangle^2}\frac{1}{2\sqrt{\pi}\sigma\nu_{c}} =\frac{1}{\sqrt{\pi}\sigma\nu_{c}},
\label{eq:ranas}
\end{equation}
where $\langle r\rangle$ and $\langle r^2\rangle$ are the first and second moments of the exponential distribution used to generate the random numbers in Eq.~\eqref{eq:rancom}, and their ratio
\begin{equation}
\frac{\langle r^2\rangle}{\langle r\rangle^2}=2 
\end{equation}
is determined by the fluctuations of the energy components of the LDoS of the random state with respect to the exact Gaussian envelope for state IV. Interestingly, one sees that due to this ratio, the value of $S_{P}^{\infty}$ is twice as large as the asymptotic value of the classical survival probability, which is given by Eq.~(\ref{eqn:resSPCAsymptapprox}). If the fluctuations of the energy components of the LDoS of the random state were absent and $\langle r^2\rangle/\langle r\rangle^2$ was one, $S_{P}^{\infty}$  would coincide with the classical result. The origin of these fluctuations is rooted in a remaining structure present even in strongly chaotic eigenstates (the so-called nodal structure~ \cite{Heller1984,Heller1991,Nonnenmacher2010}), which effectively limits the ergodicity of the quantum evolution~\cite{Stechel1985} when compared to the classical dynamics.

We close this discussion with an interesting observation. The minimum value of $S_P^{(r)}(t)$, reached right after the Gaussian decay, coincides exactly with $\mathfrak{S}_{P}^\infty$. That is, before the ramp towards equilibration, $S_P^{(r)}(t)$ stabilizes at the value associated with the ergodicity of the classical evolution. Why the random states are able to reach this greatest level of spreading to later contract to the value $S_{P}^{\infty}$ of maximal quantum ergodicity is an open question to us.

%%%%%%%%%%%%%%%%%%%%%%%%%%%%%%%%%%%%%%%%%%%%%%%%%%%%%%%
\subsubsection{Quantum Asymptotic Values and Unstable Periodic Orbits}

The purpose of this subsection is to show numerically that most  initial coherent states in the chaotic region are indeed marginally affected by unstable periodic orbits, being well described by the behavior of the survival probability reported in Fig.~\ref{fig:8}. 

%%%%%%%%%%%%%%%%%%%%%%%%%%%%%%%%%%%%%%%%%%%%%%%%%%%%%%%
\begin{figure}[h!]
\centering{
\includegraphics[width=11cm]{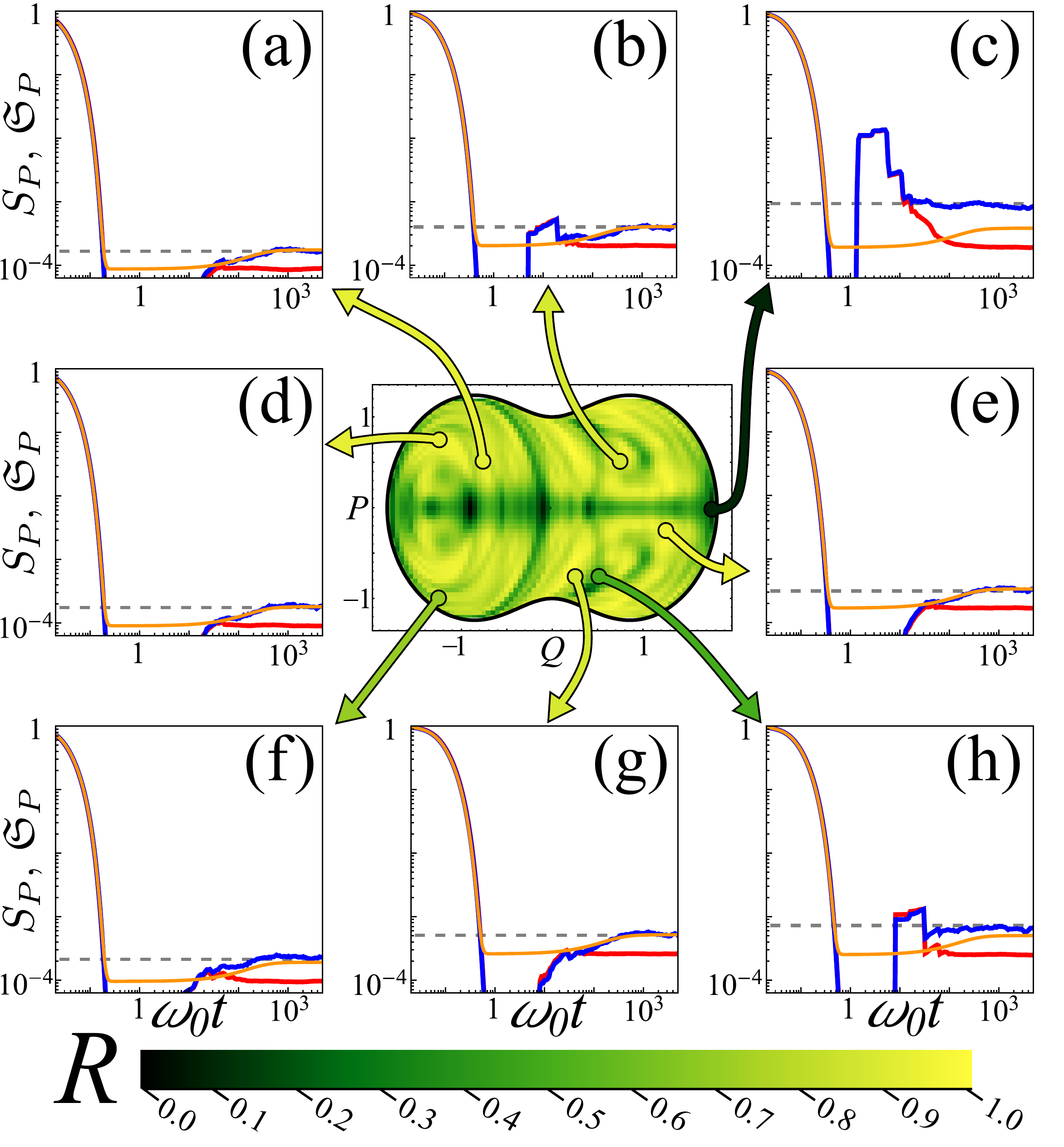}
}
\caption{Central panel: Map of the ratio $R$ [Eq.~\eqref{eq:R}] for a mesh of initial coherent states in the chaotic regime. Panels (a)-(h) show the quantum (blue solid line) and classical (red solid line) temporal averaged survival probability and the analytical expression (orange solid line) for random states [Eq.~\eqref{eqn:SPRMT}]. The horizontal dashed line represents the quantum asymptotic value $S_{P}^{\infty}$.
}
\label{fig:9}
\end{figure}
%%%%%%%%%%%%%%%%%%%%%%%%%%%%%%%%%%%%%%%%%%%%%%%%%%%%%%%

In the central panel of Fig.~\ref{fig:9}, we show the ratio
\begin{equation}
R=\frac{ S_P^{(r),\infty} }{S_P^{\infty } } =\frac{2 \, \mathfrak{S}_P^{\infty} }{S_P^{\infty } }
\label{eq:R}
\end{equation}
for a mesh of several initial coherent states distributed over the same Poincar\'e surface as in  Fig.~\ref{fig:1}~(c). In the equation above, $S_P^{(r),\infty}$ is given by the analytical expression in Eq.~\eqref{eq:ranas}, which is the asymptotic value of an initial coherent state without the influence of  unstable periodic orbits. We see that for most of the initial coherent states the ratio $R$ is very close to one (light color), which implies that $\dfrac{S_P^{\infty }}{\mathfrak{S}_P^{\infty}} \approx 2$.  There are few values of $R$ that are smaller than one (dark color), $\dfrac{S_P^{\infty }}{\mathfrak{S}_P^{\infty}} >2 $, which indicates initial states affected by unstable periodic orbits. The ratio $R$ is twice the factor $\mathcal{F}$ introduced in \cite{Heller1987} to gauge the fraction of phase space explored by the quantum state.

Around the central panel of Fig.~\ref{fig:9}, several plots of the quantum (blue solid line) and classical (red solid line) survival probabilities averaged over temporal windows are shown together with the analytical expression in Eq.~(\ref{eqn:SPRMT}) for  random initial states (orange solid line). Panels (a), (b), (d)-(g) make it clear that a ratio $R\approx 1$ $\left(\dfrac{S_P^{\infty }}{\mathfrak{S}_P^{\infty}} \approx 2\right)$ leads to a generic behavior of the quantum $S_P(t)$,  well described by $S_P^{(r)} (t)$. In all these panels, there appears a ramp towards equilibration associated with the presence of correlated eigenvalues. These random-like coherent states contrast with those where ratio $R<1$ $\left(\dfrac{S_P^{\infty }}{\mathfrak{S}_P^{\infty}} >2 \right)$, which are affected by quantum scarring, as in Fig.~\ref{fig:9}~(c) and (h). These latter cases lead to recurrences in the evolution of the survival probability and a strong influence of the unstable periodic orbits in the quantum dynamics.

%%%%%%%%%%%%%%%%%%%%%%%%%%%%%%%%%%%%%%%%%%%%%%%%%%%%%%%
%%%%%%%%%%%%%% SEC.5  CONCLUSIONS %%%%%%%%%%%%%%%%
%%%%%%%%%%%%%%%%%%%%%%%%%%%%%%%%%%%%%%%%%%%%%%%%%%%%%%%
\section{CONCLUSIONS}
\label{sec:5}

This work provides a comprehensive study of the dynamics and equilibration process of the Dicke model in the regular and chaotic regimes. The classical dynamics obtained via the truncated Wigner approximation (TWA)  explains some features of the quantum evolution. We show that the quantum and classical survival probabilities agree extremely well up to the point where one of the two saturates. After equilibration, the classical $ \mathfrak{S}_P(t)$ reaches a constant value, while the quantum $S_P(t)$ shows fluctuations due to the discreteness of the energy spectrum.

By comparing the entire quantum and classical evolutions, we identify properties other than the quantum fluctuations that are also purely quantum and analyze how they affect the equilibration times. We build a broad picture of the dynamics that counts with two general cases in the regular region and two in the chaotic region. These four scenarios are distinguished via the ratio between the asymptotic values of the quantum and classical survival probabilities, $\dfrac{ S_P^{\infty }}{   \mathfrak{S}_P^{\infty}   } $. 
We itemize below our main findings.

{\bf -- At short times, the survival probability of coherent states reaches values close to zero.} This reflects the localization of the Wigner distribution in phase space and the fact that it moves out from its starting region. Later revivals of $S_P(t)$ are associated with the return of the Wigner distribution to its starting classical region. 

{\bf -- The long-time dynamics in the regular regime depends on the proximity to a separatrix.} For an initial state close to a separatrix, dynamical tunneling between classically disconnected phase-space regions leads to an asymptotic value of the quantum survival probability slightly lower than the classical one, $\dfrac{ S_P^{\infty }}{   \mathfrak{S}_P^{\infty}   } \lesssim 1$, while away from the separatrix the asymptotic values coincide, $\dfrac{ S_P^{\infty }}{   \mathfrak{S}_P^{\infty}   } = 1$. This difference holds even though both states have the same energy.

{\bf -- The equilibration times of the classical and quantum dynamics coincide in the regular regime, but differ in the chaotic region}.  In the latter case, dynamical manifestations of quantum chaos associated with quantum scars and spectral correlations separate the curves of the classical and quantum survival probability, one saturating earlier than the other.

{\bf -- For initial coherent states in the chaotic region and affected by quantum scarring, the equilibration of the quantum dynamics happens before the classical one.} In this case, the number of energy eigenstates that participate in the quantum evolution is small. This effectively shrinks the  phase space  available for the quantum evolution as compared to what is available for the classical evolution, so quantum equilibration occurs sooner. This is reflected in the value of the ratio $\dfrac{ S_P^{\infty }}{   \mathfrak{S}_P^{\infty}   } >2$.
The larger this ratio is, the larger the degree of scarring.

{\bf -- For initial states highly delocalized in the chaotic regime (ergodic states), the equilibration of the quantum dynamics happens after the classical one.} This is due to the emergence of the correlation hole, which is caused by the correlations between the eigenvalues and is nonexistent in the classical limit.  In this case, the quantum evolution of the survival probability at long times coincides with that for random initial states. This analogy allows us to derive analytically the asymptotic value $S_P^{\infty}$ and the relaxation time. We find that $\dfrac{ S_P^{\infty }}{   \mathfrak{S}_P^{\infty}   } =2$. This value signals the onset of maximal quantum ergodicity. The fact that this ratio is not 1 indicates that the quantum evolution is more restricted than the classical one due to remaining structures of the quantum states. We emphasize that the correlation hole is a universal behavior that emerges in any chaotic quantum model, provided the initial state is non-scarred. 

Tunneling, quantum scars, and the correlation hole are purely quantum properties that are not exclusive to the Dicke model. We therefore expect our results to be applicable to other models with regular and chaotic regimes and with a quantum-classical correspondence.

%%%%%%%%%%%%%%%%%%%%%%%%%%%%%%%%%%%%%%%%%%%%%%%%%%%%%%%
%%%%%%%%%%%%%% ACKNOWLEDGEMENTS %%%%%%%%%%%%%%%%
%%%%%%%%%%%%%%%%%%%%%%%%%%%%%%%%%%%%%%%%%%%%%%%%%%%%%%%         
\section*{ACKNOWLEDGEMENTS}

We acknowledge the support of the Computation Center - ICN, in particular to Enrique Palacios, Luciano D\'iaz, and Eduardo Murrieta, and valuable conversations with Jonathan Torres and Jorge Ch\'avez-Carlos. We acknowledge financial support from Mexican CONACyT project CB2015-01/255702, DGAPA- UNAM projects IN109417 and IN104020. LFS is supported by the NSF grant No. DMR-1936006. LFS and JGH thank the hospitality of the 
Aspen Center for Physics and the Simons Center for Geometry and Physics at Stony Brook University, where some of the research for this paper was performed.

%%%%%%%%%%%%%%%%%%%%%%%%%%%%%%%%%%%%%%%%%%%%%%%%%%%%%%%
%%%%%%%%%%%%%% APPENDIX %%%%%%%%%%%%%%%%
%%%%%%%%%%%%%%%%%%%%%%%%%%%%%%%%%%%%%%%%%%%%%%%%%%%%%%%

\appendix

%%%%%%%%%%%%%%%%%%%%%%%%%%%%%%%%%%%%%%%%%%%%%%%%%%%%%%%
%%%%%%%%%%%%%%%%%%%%%%%%%%%%%%%%%%%%%%%%%%%%%%%%%%%%%%%
\section{Classical Limit of the Dicke Hamiltonian}
\label{app:1}

To obtain the classical Hamiltonian in Eq.~(\ref{eqn:cla_hamiltonain_QP}), we use
the Glauber [Eq.~\eqref{eqn:glauber}] and Bloch [Eq.~\eqref{eqn:bloch}] coherent states expressed in terms of the general parameters $\alpha,z\in\mathbb{C}$ as a tensor product of the form,
\begin{align}
\label{eqn:tensor_glauberbloch}
|\alpha,z\rangle=& |\alpha\rangle\otimes|z\rangle
\nonumber
\\
=& \frac{e^{-|\alpha|^{2}/2}}{(1+|z|^{2})^{j}}\sum_{n=0}^{\infty}\sum_{m=-j}^{j}\sqrt{\left(\begin{array}{c}2j\\ j+m\end{array}\right)}\frac{\alpha^{n}z^{j+m}}{\sqrt{n!}}|n\rangle\otimes|j,m\rangle,
\end{align}
and take the expectation value of the Dicke Hamiltonian $\hat{H}_{D}$~\cite{Deaguiar1991,%%*
Deaguiar1992}, 
\begin{align}
\label{eqn:cla_hamiltonian}
H_{cl}=& \langle\alpha,z|\hat{H}_{D}|\alpha,z\rangle
\nonumber
\\
=& \omega|\alpha|^{2}-j\omega_0\frac{1-|z|^{2}}{1+|z|^{2}}+{\gamma}\sqrt{2j}\frac{z+z^{\ast}}{1+|z|^{2}}(\alpha+\alpha^{\ast}).
\end{align}
Considering the harmonic oscillator $\alpha=\sqrt{\dfrac{j}{2}}(q+i p)$ and the Bloch sphere $z=\sqrt{\dfrac{1+j_{z}}{1-j_{z}}}e^{-i\phi}$ parameters in terms of canonical variables $(q,p)$ and $(\phi,j_{z})$, we obtain
\begin{equation}
\label{eqn:cla_hamiltonian_phijz}
H_{cl}=j\left[\frac{\omega}{2}(q^{2}+p^{2})+\omega_0j_{z}+2\gamma\sqrt{1-j_{z}^{2}}q\cos(\phi)\right].
\end{equation}
To finally get Eq.~(\ref{eqn:cla_hamiltonain_QP}), we perform
 a canonical transformation to the atomic variables  $(j_{z},\phi)\rightarrow (Q,P)$, where $Q=\sqrt{2(1+j_{z})}\cos(\phi)$ and $P=-\sqrt{2(1+j_{z})}\sin(\phi)$ satisfy the Poisson bracket $\{Q,P\}=1$, and rescale the overall classical Hamiltonian $H_{cl}$ by $j$. 

To obtain the representation of coherent states given by Eq.~\eqref{eqn:glauber} and Eq.~\eqref{eqn:bloch}, we express the coherent state parameters $\alpha$ and $z$ in terms of the canonical variables $(q,p,Q,P)$, such that $\alpha=\sqrt{\dfrac{j}{2}}(q+i p)$ and $z=\dfrac{1}{\sqrt{4-Z^{2}}}(Q+iP)$ with $Z^{2}=Q^{2}+P^{2}$.

\begin{figure}[h!]
\centering{
\includegraphics[width=10cm]{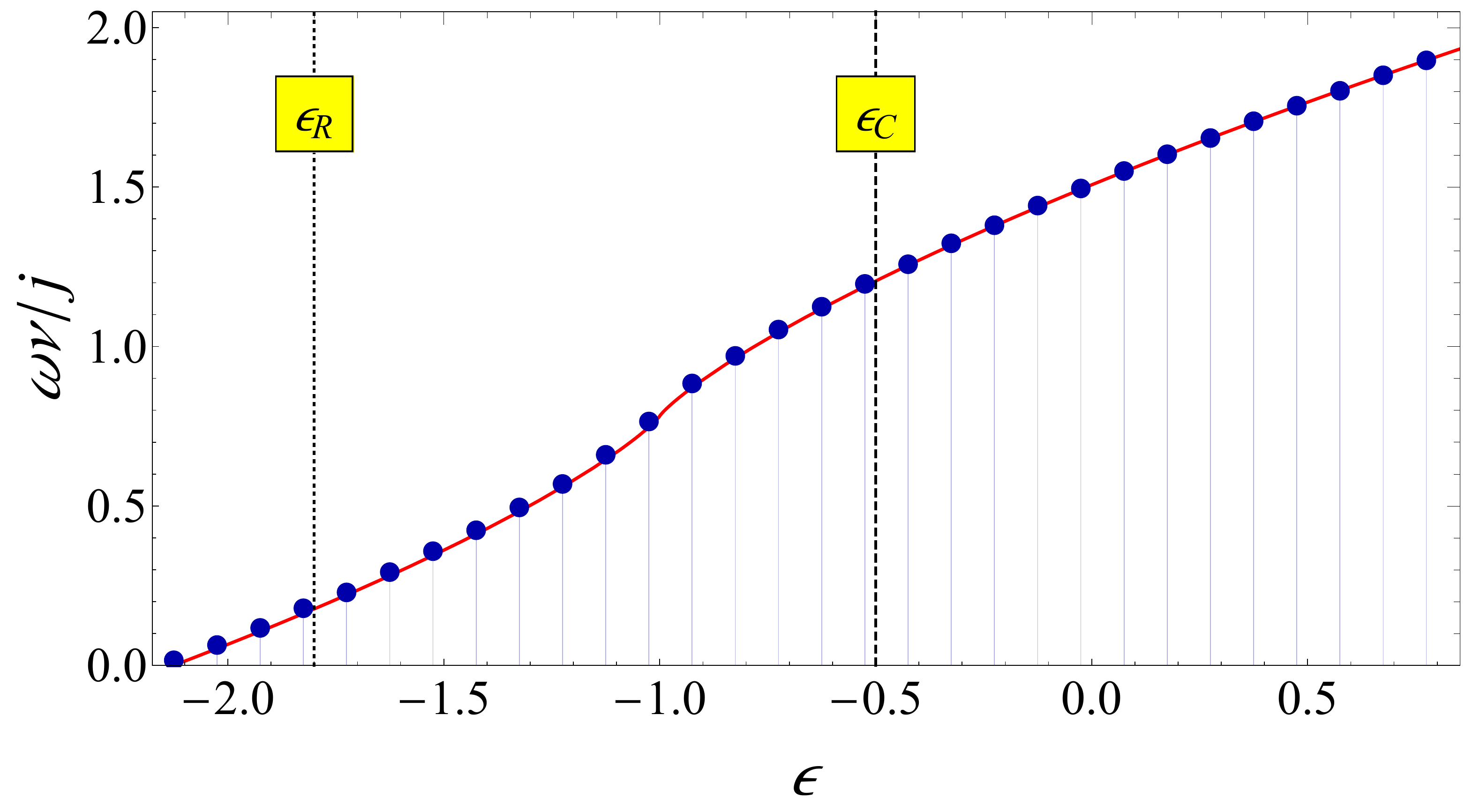}
}
\caption{Density of states obtained numerically (blue dots) with bin size $\Delta\epsilon=0.1$ and analytical expression~\eqref{eqn:DoS} (red solid line). Hamiltonian parameters: $\omega=\omega_{0}$, $\gamma=2\gamma_c$, and $j=100$. Vertical black dotted and dashed lines indicate the energies $\epsilon_{R}=-1.8$ and $\epsilon_{C}=-0.5$ selected for our studies, respectively. A truncated Hilbert space was employed, ensuring $N_{c}=30825$ converged eigenstates and eigenenergies, which range from the ground state energy $\epsilon_{GS}=-2.125$ up to a truncation energy $\epsilon_{T}=0.853$.
}
\label{fig:10}
\end{figure}

With the classical Hamiltonian $H_{cl}$, a semi-classical approximation to the DoS can be found by integrating the phase space volume available for a given energy. It reads~\cite{Brandes2013,Bastarrachea2014a}
\begin{equation}
\label{eqn:DoS}
\nu( \epsilon)=\frac{2j}{\omega}\left\{\begin{array}{l}\frac{1}{\pi}\int_{y_{-}}^{y_{+}}\dif y\cos^{-1}\left(\sqrt{\frac{2(y-\epsilon)}{\bar{\gamma}^{2}(1-y^{2})}}\right), \epsilon_{gs}\leq\epsilon<-1  \\ 
\frac{1+\epsilon}{2}+\frac{1}{\pi}\int_{\epsilon}^{y_{+}}\dif y\cos^{-1}\left(\sqrt{\frac{2(y-\epsilon)}{\bar{\gamma}^{2}(1-y^{2})}}\right), |\epsilon|\leq 1 \\ 
1, \epsilon>1\end{array}\right.,
\end{equation}
where $y_{\pm}=-\bar{\gamma}^{-1}\left(\bar{\gamma}^{-1}\mp\sqrt{2(\epsilon-\epsilon_{0})}\right)$ and $\bar{\gamma}=\gamma/\gamma_{c}$. The ground state energy is given by $\epsilon_{gs}=-1$  for the normal phase, and by $\epsilon_{gs}=\epsilon_{0}=-\dfrac{1}{2}\left(\bar{\gamma}^{2}+\bar{\gamma}^{-2}\right)$ for the superradiant phase. In Fig.~\ref{fig:10}, we compare the DoS obtained numerically with Eq.~\eqref{eqn:DoS}, showing excellent agreement. The figure also shows the energies selected  for our studies throughout this paper.

%%%%%%%%%%%%%%%%%%%%%%%%%%%%%%%%%%%%%%%%%%%%%%%%%%%%%%%
%%%%%%%%%%%%%%%%%%%%%%%%%%%%%%%%%%%%%%%%%%%%%%%%%%%%%%%

%%%%%%%%%%%%%%%%%%%%%%%%%%%%%%%%%%%%%%%%%%%%%%%%%%%%%%%
%%%%%%%%%%%%%%%%%%%%%%%%%%%%%%%%%%%%%%%%%%%%%%%%%%%%%%%
\section{Wigner Function of Glauber and Bloch Coherent States}
\label{app:3}

As we see from \ref{app:1}, the coherent states for the Dicke model are the product of Glauber and Bloch coherent states [Eq.~\eqref{eqn:tensor_glauberbloch}]. Thus, the associated Wigner function is the product of the Wigner functions associated to the Glauber and Bloch states. For a Glauber state $|\alpha\rangle$ with $\alpha(q,p)=\sqrt{\frac{j}{2}}(q+i p)$, the Wigner function is given by a normal distribution
\begin{equation} 
\label{eqn:Wqp_def}
w_{q_0,p_0}(q,p)=\frac{j}{\pi}e^{-j d^2},
\end{equation}
with $d=\sqrt{\left(q-q_0 \right)^2 + \left(p-p_0 \right)^2} $. 
For a Bloch coherent state, the Wigner function  in variables $(\theta,\phi)$ may be written as a sum of Legendre polynomials $P_k(x)$~\cite{Klimov2019}
\begin{equation} 
w_{\theta_0,\phi_0}(\theta,\phi)=\frac{(2j)!}{4\pi}\sum_{k=0}^{2j}\sqrt{\frac{(2k+1)}{(2j-k)!(2j+k+1)!}} P_k (\cos \Theta), 
\end{equation}
where $\Theta$ is the angle between $(\theta,\phi)$ and $(\theta_0,\phi_0)$ obtained from
\begin{equation}
\cos\Theta = \cos \theta \cos \theta_0 + \sin \theta \sin \theta_0 \cos(\phi - \phi_0).
\end{equation}
For large $j$ values this is very well approximated by a normal distribution on the Bloch sphere
\begin{equation} 
\label{eqn:WQP_def}w_{\theta_0,\phi_0}(\theta,\phi)\approx\frac{j}{\pi}e^{-j \Theta^2}.
\end{equation}
This approximation is very useful, because sampling from normal distributions is easy and computationally cheap. The complete coherent Wigner function is just the product of the Wigner functions given by Eq.~\eqref{eqn:Wqp_def} and Eq.~\eqref{eqn:WQP_def},
\begin{equation} 
\label{eqn:coherentWUsed}
w_{\bm{u_0}}(\bm u)=\left (\frac{j}{\pi}\right )^2e^{-j\left(d^2 +\Theta^2  \right)}.
\end{equation}

%%%%%%%%%%%%%%%%%%%%%%%%%%%%%%%%%%%%%%%%%%%%%%%%%%%%%%%
%%%%%%%%%%%%%%%%%%%%%%%%%%%%%%%%%%%%%%%%%%%%%%%%%%%%%%%
\section{Truncated Wigner Aproximation and Monte Carlo Method}
\label{app:4}

The temporal evolution of the Wigner function is governed by the so-called Moyal equation\footnote{Formally, the Hamiltonian that should be used in the Moyal equation is the Weyl transform of the quantum Hamiltonian. Our Hamiltonian was not obtained by Weyl transformation but by the calculation of the expectation value with coherent states. The difference between both Hamiltonians turns out to be equal to the constant energy $\frac{\omega + \omega_0}{2}$. Because only the derivatives of the Hamiltonian appear in the equation, this constant number makes no actual difference.}
\cite{Klimov2017}
\begin{equation}
\frac{\partial w}{\partial t}(\bm{u},t) = \left \{ w(\bm u,t) ,h_{cl}(\bm u)\right\}_M.
\end{equation}
Here, $\left \{A,B \right \}_M=\dfrac{2}{\hbar} A\sin\left[\dfrac{\hbar}{2} \left(\overleftarrow{\partial_{\bm{q}}} \overrightarrow{\partial_{\bm{p}}} - \overleftarrow{\partial_{\bm{p}}} \overrightarrow{\partial_{\bm{q}}}  \right) \right]B$ represents the Moyal bracket. Taylor expanding the sine, one may write $\left \{A,B \right \}_M=\left \{A,B \right \} + \mathcal{O}\left(\hbar^2\right)$ where  $\left \{\cdot,\cdot \right \}$ is the Poisson bracket. So
\begin{equation}
\frac{\partial w}{\partial t}(\bm{u},t) = \left \{ w(\bm u,t) ,h_{cl}(\bm u)\right \} +\mathcal{O}\left(j^{-2} \right). 
\end{equation}

If we ignore the $j^{-2}$ order terms in this equation, we are left with the classical Liouville equation. This is known as the truncated Wigner approximation (TWA) and yields the correct quantum evolution for small times. 

Within this approximation, the Wigner function remains constant along classical trajectories in phase space, so the time dependence of $w(\boldsymbol u,t)$ may be written in terms of the Hamiltonian flow $\flow{t}:\mathcal{M} \to \mathcal{M}$. This function describes the time evolution of an initial condition $\boldsymbol {u_0}\in \mathcal{M}$ by
\begin{equation}
\boldsymbol u(t)=\flow{t}(\boldsymbol {u_0}),
\end{equation}
and it satisfies the one-parameter group identities $\flow{0} = \text{Id}$, $\flow{-t} =(\flow{t})^{-1}$, and ${\flow{t_1+t_2}=\flow{t_2}\circ \flow{t_1}}$. Staying constant along classical trajectories means that for any pair of times $t_1$ and $t_2$
\begin{equation}
w(\flow{t_1}(\boldsymbol u),t_1) =w(\flow{t_2}(\boldsymbol u),t_2).
\end{equation}
In particular, taking $t_2=0$ and performing an adequate change of variables,
\begin{equation}
\label{eqn:Wigner_flow}
w(\boldsymbol u,t) =w(\flow{-t}(\boldsymbol u),0). 
\end{equation}
Inserting Eq.~\eqref{eqn:Wigner_flow} into Eq.~\eqref{eqn:SPWigner}, we find Eq.~\eqref{eqn:SPconflujo}. 

Note that Eq.~\eqref{eqn:SPconflujo} may be written as
\begin{equation}
\mathfrak{S}_{P}(t)  =\left\langle  w \left(\flow{-t}(\boldsymbol u)\right) \right\rangle_{w}.
\end{equation}
Provided that $w$ is everywhere positive, as it is for coherent states [see Eq.~\eqref{eqn:coherentWUsed}], this expected value may be efficiently approximated using a Monte Carlo method,
\begin{equation} 
\mathfrak{S}_{P}(t)\approx  \frac{1}{M}\sum_{i=1}^M w(\flow{-t}(\boldsymbol {u_i})),
\end{equation}
where the points $\boldsymbol {u_i} \in \mathcal{M}$ are randomly sampled from the initial distribution $w$, and $M$ is a sufficiently large, albeit computationally accessible, integer.

For the plots of $\mathfrak{S}_{P}(t)$ shown in Figs. \ref{fig:4}, \ref{fig:5}, \ref{fig:7}, and \ref{fig:8}, and for the animations available in \cite{SM}, we use a value of $M=10^7$ to significantly reduce numerical noise. For Figs. \ref{fig:6} and \ref{fig:9}, because temporal averages reduce the numerical error, we only need to use $M=3\times 10^5$. 

%%%%%%%%%%%%%%%%%%%%%%%%%%%%%%%%%%%%%%%%%%%%%%%%%%%%%%%
%%%%%%%%%%%%%%%%%%%%%%%%%%%%%%%%%%%%%%%%%%%%%%%%%%%%%%%
\section{Intuitive Interpretation of the Classical Survival Probability}
\label{app:6}

An intuitive understanding of the classical survival probability may be obtained by considering a simple measurable set $\mathcal{S}\subseteq \mathcal{M}$ (a sphere, for example). Let $\boldsymbol {1}_{\mathcal{S}}$ be its indicator function\footnote{For any set $\mathcal{X}  \subseteq \mathcal{M}$, its indicator function $\bm{1}_{\mathcal{X}}:\mathcal{M} \to \left \{0,1 \right \}$ equals 1 for
  points in $\mathcal{X}$ and 0 for points outside.}. Consider the normalized distribution $\rho_\mathcal{S}=\bm{1}_{\mathcal{S}}/V_\mathcal{S}$ in lieu of the Wigner function. Here $V_\mathcal{S}=\int  \dif \bm u \, \bm{1}_{\mathcal{S}}(\bm{u})$ is the volume of $\mathcal{S}$. In this case,  to retain normalization, our prefactor becomes $1/V_\mathcal{S}$ instead of $(2\pi/j)^2$, and then, from Eq.~\eqref{eqn:SPconflujo},
\begin{equation} 
\label{eqn:SPesfera}
\mathfrak{S}_{P}(t)=\frac{1}{V_\mathcal{S}}\int_\mathcal{M} \dif \bm{u}\,  \boldsymbol {1}_{\mathcal{S}}(\bm{u})\boldsymbol {1}_{\mathcal{S}}(\flow{-t}(\bm{u})).
\end{equation}
Since $\bm{1}_{\mathcal{S}}(\flow{-t}(\bm{u}))=\bm{1}_{\flow{t}(\mathcal{S})}(\bm{u})$ and using that the product of the indicator functions of two sets is the indicator function of their intersection, we get 
\begin{equation} 
\mathfrak{S}_{P}(t)=\frac{1}{V_\mathcal{S}}\int_\mathcal{M} \dif \bm{u}\,  \boldsymbol {1}_{\mathcal{S}\cap \flow{t}(\mathcal{S})}(\bm u) = \frac{V_{\mathcal{S}\cap \flow{t}(\mathcal{S})}}{V_\mathcal{S}}.
\end{equation}
This result is easy to interpret: $\mathfrak{S}_{P}(t)$ is the percentage of the volume of $\mathcal{S}$ that is found back inside $\mathcal{S}$ after time $t$. In other words, it is the probability that a point sampled from $\mathcal{S}$ is back in $\mathcal{S}$ after time $t$~\cite{Pechukas1982,Gorin2006}.

%%%%%%%%%%%%%%%%%%%%%%%%%%%%%%%%%%%%%%%%%%%%%%%%%%%%%%%
%%%%%%%%%%%%%%%%%%%%%%%%%%%%%%%%%%%%%%%%%%%%%%%%%%%%%%%
\section{Asymptotic Value of the Classical Survival Probability}
\label{app:5}

To obtain an expression for $\mathfrak{S}_{P}^\infty$, we assume ergodicity. This is a reasonable assumption  for chaotic behaviors. If the Hamiltonian flow $\flow{t}$ is ergodic over the energy shells in phase space, then, for any real function $f(\bm u)$ of phase space, temporal averages in composition with the flow are equal to space averages over the corresponding energy shell, that is, for any fixed point $\bm u \in \mathcal{M}$ with energy $E=H_{cl}(\bm{u})$,
\begin{align}
\expval{f(\flow{t}(\bm u))}_{t\to\infty}=& \expval{f}_{E}
\nonumber
\\
=& \frac{j^2}{(2\pi)^2 \nu(E)} \int_\mathcal{M} \dif \bm{v} \, \delta(H_{cl}(\bm{v}) - E) f(\bm v),
\end{align}
where $j^{-2}(2\pi)^2 \nu(E)$ is the volume of the energy shell for $E/j$ in $\mathcal{M}$, and $\nu(E)$ is given by Eq.~\eqref{eqn:DoS}.

It is then straightforward to calculate
\begin{align}
\expval{\mathfrak{S}_{P}(t)}_{t\to\infty}=& \left(\frac{2\pi}{j} \right)^2 \int_{\mathcal{M}} \dif \bm{u}\,  w(\bm u) \expval{w(\flow{-t}(\bm u))}_{t\to\infty}
\nonumber
\\
=& \left(\frac{2\pi}{j} \right)^2 \int_{\mathcal{M}} \dif \bm{u}\,  w(\bm u) \expval{w}_{H_{cl}(\bm{u})}
\nonumber
\\
=& \left(\frac{2\pi}{j} \right)^4 \int_{E_{gs}}^\infty \dif E\, \expval{w}_{E}^2 \nu(E).
\end{align}
The last equality is obtained by using ${1=\int_{E_{gs}}^\infty \dif E\,  \delta(E-H_{cl}(\bm{u}))}$ inside the integral, a change in the integration order, and a substitution of the value of $\expval{w}_E$. In the above, $E_{gs}$ represents the ground state energy. The classical energy distribution for the state associated to $w$ is 
\begin{align}
\rho_{cl}(E)= & \int_\mathcal{M}{\dif \bm{u}\,  \delta(E-H_{cl}(\bm{u})) w(\bm{u})}
\nonumber
\\
=&(2\pi)^2 j^{-2} \nu(E) \expval{w}_{E},
\end{align}
therefore
\begin{equation}
\expval{\mathfrak{S}_{P}(t)}_{t\to\infty}= \int_{E_{gs}}^\infty \dif E\, \frac{\rho_{cl}^{2}(E)}{\nu(E)}.
\end{equation}
This result is exact, but we may approximate $\nu(E)$ [see Eq.~\eqref{eqn:DoS}] by the value at the center of the classical energy distribution ($\nu(E)\approx\nu(E_{c})=\nu_{c}$) to get
\begin{equation}
\label{eqn:resintegral}
\expval{\mathfrak{S}_{P}(t)}_{t\to\infty}= \frac{1}{\nu_{c}}\int_{E_{gs}}^\infty \dif E\, \rho_{cl}^{2}(E).
\end{equation}
By further approximating $\rho_{cl}$ with the Gaussian distribution given by Eq.~\eqref{eqn:gaussian_LDoS}, we obtain Eq.~\eqref{eqn:resSPCAsymptapprox}.

%%%%%%%%%%%%%%%%%%%%%%%%%%%%%%%%%%%%%%%%%%%%%%%%%%%%%%%
%%%%%%%%%%%%%%%%%%%%%%%%%%%%%%%%%%%%%%%%%%%%%%%%%%%%%%%
\section{Analytical Expression of the Survival Probability for a Random Ensemble with a Gaussian Energy Profile}
\label{app:2}

An analytical expression for the survival probability averaged over an ensemble of random initial states was derived in ~\cite{Lerma2019}. The properties of the ensemble are as follows. Its members are constrained to have a smooth LDoS $\rho(E)$  and their energy components are given by
\begin{equation}
|c_{k}^{(r)}|^2=\frac{r_{k}\,\rho(E_{k})}{\mathcal{A}\,\nu(E_{k})},
\end{equation}
where $\nu(E)$ is the DoS, the numbers $r_k$ are randomly  generated from a given probability distribution $P(r)$, and  $\mathcal{A}=\sum_{q}r_{q}\rho(E_{q})/\nu(E_{q})$ is a normalization constant. This gives,
\begin{equation}
S_{P}^{(r)}(t)=\frac{1- S_{P}^{(r),\infty} }{\eta-1}\left[\eta S_{P}^{\text{bc}}(t)-b_{2}\left(\frac{Dt}{2\pi}\right)\right]+   S_{P}^{(r),\infty},
\label{eq:SPran}
\end{equation}
where the initial decay of the survival probability is dictated by 
\begin{equation}
S_{P}^{\text{bc}}(t)=\left|\int \dif E\rho(E)e^{-i E t}\right|^{2},
\end{equation}
the effective dimension of the ensemble 
\begin{equation}
\eta=\frac{1}{\int dE \rho^2(E)/\nu(E)}\approx\frac{\nu_{c}}{\int \dif E\,\rho^{2}(E)},
\end{equation}
where $\nu_c=\nu(E_c)$ is the density of states evaluated in the center of the energy profile, 
and the asymptotic value of the survival probability corresponds to
\begin{equation}
S_{P}^{(r),\infty}=
\frac{\langle r^{2}\rangle}{\langle r\rangle^{2}}\frac{1}{\eta},\\ 
\end{equation}
with $\langle r^n\rangle$  the $n$-th moments of the distribution $P(r)$.

The function $b_2$ is the two-level form factor of the GOE  \cite{MehtaBook} 
\begin{equation}
\label{eqn10}
\begin{split}
b_{2}(\bar{t})=  [1-2\bar{t}+\bar{t}\ln(2\bar{t}+1)]\Theta(1-\bar{t})  +\left[\bar{t}\ln\left(\frac{2\bar{t}+1}{2\bar{t}-1}\right)-1\right]\Theta(\bar{t}-1),
\end{split}
\end{equation}
where $\Theta$ is the Heaviside step function. The factor $D$ in the argument of $b_2$ is the mean level spacing of the correlated eigenvalues.

For the random ensemble that we consider in this paper, $\rho(E)$ has a Gaussian energy profile  [see Eq.~\eqref{eqn:gaussian_LDoS}] and the random numbers $r_k$ are generated from  the exponential distribution $P(r)=\lambda e^{-\lambda r}$ [see Fig.~\ref{fig:3}(m)] with $\langle r^{n}\rangle=n!/\lambda^{n}$, which implies that 
\begin{align}
S_{P}^{\text{bc}}(t)&=e^{-\sigma^{2}t^{2}}, \\
\eta &= 2\sqrt{\pi}\sigma\nu_{c}, \\
S_{P}^{(r),\infty}&=\frac{2}{\eta}=\frac{1}{\sqrt{\pi}\sigma\nu_{c}}.
\end{align}

Because the correlations in the spectrum appears only for energy levels in the same parity sector, the mean level spacing of correlated eigenvalues  is  $D=(D_++D_-)/2$, where  the mean level spacing for each parity sector,  $D_\pm=1/\nu_{\pm}$,   is given by the respective density of states. These densities are, in turn, given by $\nu_\pm=\nu_c/2$, with $\nu_c$ the density of states of the whole spectrum, yielding   $D=2/\nu_c$. From the previous  results for $S_{P}^{(r),\infty}$, $\eta$, $S_{P}^{\text{bc}}$, and $D$, we get Eq.~\eqref{eqn:SPRMT}.

To derive the relaxation time $t_r$ of the ensemble-averaged survival probability,   we consider the asymptotic form of  $b_2$, which grows toward saturation following a power-law behavior
\begin{equation}
b_2\left( \frac{t}{\pi \nu_c } \right) \rightarrow  \frac{\pi^2 \nu_c^2}{12 t^2} \hspace{0.4 cm } \text{for} \hspace{0.4 cm } \frac{t}{\pi\nu_c} \gg 1 .
\label{Eq:longB2}
\end{equation}
At this temporal scale, the contribution of the initial decay  $S_{P}^{\text{bc}}$ is negligible and the asymptotic form of \eqref{eq:SPran} is given by 
\begin{equation}
S_{P}^{(r)}(t)\rightarrow -\frac{1}{\eta}b_2\left(\frac{t}{\pi \nu_c}\right)+ S_{P}^{(r),\infty} = S_{P}^{(r),\infty}\left(1-\frac{\pi^2 \nu_c^2}{24 t^2}\right), 
\label{eq:spasym}
\end{equation}
where in the last step we have used $\eta=2/S_{P}^{(r),\infty}$.  We define the relaxation time according to
\begin{equation}
\label{eq:spasym2}
 S_{P}^{(r)}(t_r)  = (1-\delta_{S_P})  S_{P}^{(r),\infty} ,
\end{equation}
where $\delta_{S_P}$ is a small parameter determining the point where $ S_{P}^{(r)}(t)$ is already within the fluctuations around the asymptotic value. By substituting Eq.~\eqref{eq:spasym} in  Eq.~\eqref{eq:spasym2}, we obtain Eq.~\eqref{eq:eqtime}.

\section*{References}
%%\bibliography{BIB_ChCs_02}

\providecommand{\newblock}{}

\end{document}